\documentclass[aps,pre,twocolumn,showpacs]{revtex4}

\usepackage{graphicx}
\usepackage{amsmath,amscd,amssymb}
\usepackage{color}

\newcommand{\be}{\begin{equation}}
\newcommand{\ee}{\end{equation}}
\newcommand{\bea}{\begin{eqnarray}}
\newcommand{\eea}{\end{eqnarray}}
\newcommand{\lan}{\left\langle}
\newcommand{\ran}{\right\rangle}
\newcommand{\br}{\mathbf{r}}

\begin{document}

\title{Ionic exclusion phase transition in neutral and weakly charged cylindrical nanopores}

\author{Sahin Buyukdagli\footnote{Email:~\texttt{buyuk@irsamc.ups-tlse.fr}}, Manoel Manghi\footnote{Email: \texttt{manghi@irsamc.ups-tlse.fr}}, and John Palmeri\footnote{Email: \texttt{john.palmeri@irsamc.ups-tlse.fr}}}
\affiliation{Universit\'e de Toulouse; UPS; \\ Laboratoire de Physique Th\'eorique (IRSAMC); F-31062 Toulouse, France}
\affiliation{CNRS; LPT (IRSAMC); F-31062 Toulouse, France}
\date{\today}

\begin{abstract}
A field theoretic variational approach is introduced to study ion penetration into water-filled cylindrical nanopores in equilibrium with a bulk reservoir \textcolor{black}{(S. Buyukdagli, M. Manghi, and J. Palmeri, Phys. Rev. Lett., \textbf{105}, 158103 (2010))}. It is shown that  an ion located in a neutral pore undergoes two opposing mechanisms: (i) a deformation of its surrounding ionic cloud of opposite charge, with respect to the reservoir, which increases the surface tension and tends to exclude ions form the pore, and (ii) an attractive contribution to the ion self-energy due to the increased screening with ion penetration of the repulsive image forces associated with the dielectric jump between the solvent and the pore wall. For pore radii around 1~nm and bulk concentrations lower than 0.2~mol/L, this mechanism leads to a first-order phase transition, similar to capillary ``evaporation'', from an ionic-penetration state to an ionic-exclusion state. The discontinuous phase transition exists within the biological concentration range ($\sim0.15$~mol/L) for small enough membrane dielectric permittivities ($\epsilon_{\rm m}<5$). In the case of a weakly charged pore, counterion penetration exhibits a non-monotonic behavior and is characterized by two regimes: at low reservoir concentration or small pore radii, coions are excluded and  counterions enter the pore due enforce electroneutrality; dielectric repulsion (image forces) remain strong and the counterion partition coefficient decreases with increasing reservoir concentration up to a characteristic value; for larger reservoir concentrations, image forces are screened and the partition coefficient of counterions increases with the reservoir electrolyte concentration, as in the neutral pore case. Large surface charge densities ($>2\times10^{-3}$~e/nm$^2$) suppress the discontinuous transition by reducing the energy barrier for ion penetration and shifting the critical point towards very small pore sizes and molar reservoir concentrations. Our variational method is also compared to a previous self-consistent approach and yields important quantitative corrections. The role of the curvature of dielectric interfaces is highlighted by comparing ionic penetration into slit and cylindrical pores. Finally, a charge regulation model is introduced in order to explain the key effect of $p$H on ionic exclusion and explain the origin of observed time-dependent nanopore electric conductivity fluctuations and their correlation with those of the pore surface charge.

\end{abstract}

\maketitle

\section{Introduction}

Electrostatic forces induced by macroscopic dielectric bodies immersed in water regulate important phenomena such as the stability of colloidal suspensions~\cite{Holm}, membrane assemblies~\cite{Dub}, and ion selectivity by synthetic membranes~\cite{Synt} as well as in biological nanopores~\cite{Bont}. The image forces induced by the dielectric permittivity jump between dielectric bodies and the solvent surrounding them play a central role in these phenomena. Indeed, the equilibrium  of similarly charged objects in a solvent is driven by the competition between the repulsive electrostatic interaction of their charges and the attractive van der Waals forces that originate from their low dielectric permittivity. This picture, valid at low ionic concentrations, is also the basis of the DLVO theory~\cite{DLVO}. A very similar competition is known to determine the permeability of biological channels that regulate ion exchange between the exterior and interior of cells. The strong dielectric discontinuity between the membrane (dielectric permittivity $\epsilon_{\rm m}\simeq2$) and the water-filled channel ($\epsilon_{\rm w}\simeq78$) is at the origin of the high potential barrier for ion penetration into the pore. In 1969, Parsegian found that the energetic cost to move an ion from the bulk reservoir into a cylindrical pore of infinite length and radius $a=0.2$~nm is approximately 16 $k_BT$~\cite{Pars}. This result clearly suggests that at room temperature, any biological pore would be totally impermeable to ions. From the numerical solution of the Debye-H\"{u}ckel (DH) equation, it was later shown that the consideration of the finite length of the channel reduces this barrier up to 6 $k_BT$~\cite{NSDH} for short enough pores. Levin~\cite{Levin} recently proposed an approximative, but reasonably  accurate, analytical solution of the DH equation for a finite length cylinder. He also showed that the introduction of surface charges can further reduce this energy barrier by attracting counterions electrostatically. Several perturbative approaches around weak-coupling (WC)  theory [mean-field (MF) theory corrected up to one-loop or Debye-Huckel (DH) order] or  strong coupling (SC) theory (virial expansion)~\cite{ShlovsSC,NetzSC} have been  developed for charged slit pores~\cite{Attard,Zeks,netzCOR,PodKan,DeanHor}, concentric cylinders~\cite{DavidCYL} as well as for two like-charged dielectric cylindrical bodies~\cite{NajiKand}. Because these methods tend to misevaluate ionic correlations, especially in the presence of strong dielectric discontinuities, it becomes imperative to develop analytical or numerical tools valid over a larger parameter range.

\textcolor{black}{
In the case of ions confined into a closed geometry characterized by a dielectric discontinuity, the existence of an infinite number of image charges significantly complicates reliable numerical calculations such as MC simulations~\cite{netzREV}. Although new MC algorithms for such systems have been developed and applied to planar dielectric slabs~\cite{Arnold1,Arnold2,Jho1,Jho2,Jho3}, at the present, we are not aware of any application of these methods to dielectric cylinders where the curvature of the interface is likely to cause a further complication of the numerical task. When MC data for ionic penetration into dielectric cylinders become available in the future, it will be interesting to test the validity of approximative theoretical methods, such as the one presented in this article.
}

Differential equations for electrostatic potentials, that partially take into account ionic correlations at a non-linear level (i.e. beyond the DH theory) were derived by Netz and Orland~\cite{netz} within a field-theoretic variational formalism. However these variational equations are too complicated to be solved analytically or even numerically~\cite{netz}. A restricted tractable variational method for neutral single and double planar interface as well as for spherical systems was proposed in Ref.~\cite{curtis,hatlo,hatlo_review}. Because of the piecewise inverse variational screening length introduced in~\cite{hatlo}, the numerical minimization procedure is technically involved. We have recently proposed a simpler variational method to investigate the case of charged planar interfaces and slit pores~\cite{Nous}. The approach is based on a generalized Onsager-Samaras approximation~\cite{onsager_samaras} which assumes an electrostatic kernel with a uniform variational inverse screening length $\kappa_v$ that may differ from the bulk value. In this article, we extend this variational method to the case of neutral and charged cylindrical pores~\cite{prl}, by using a constant variational Donnan potential $\phi_0$ which enforces electroneutrality in the charged pore~\cite{Nous}. We show here that the extremization of the variational grand potential with respect to $\kappa_v$ and $\phi_0$ yields two coupled variational equations,
\bea
\label{VarEq1}
&&\kappa_v^2=\frac{4\pi\ell_B}{\lan\frac{\partial w(\br;\kappa_v)}{\partial \kappa_v}\ran}\sum_iq_i^2\rho_{b,i}
\lan e^{-\Phi_i(\br;\kappa_v)} \frac{\partial w(\br;\kappa_v)}{\partial \kappa_v}\ran\\
\label{VarEq2}
&&\int_{S_{\rm p}}\mathrm{d}\mathbf{S}\;\sigma_s=V_{\rm p}\sum_iq_i\rho_{b,i}\lan e^{-\Phi_i(\br;\kappa_v)}\ran,
\eea
where  we define the pore average as $\lan A(\br) \ran\equiv \int_{V_{\rm p}}\frac{\mathrm{d}\br}{V_{\rm p}} A(\br)$, $V_{\rm p}$ (p for pore) stands for the volume occupied by the confined ions, $S_{\rm p}$ is the charged surface, $\sigma_s$ is the uniform surface charge density (expressed in units of the elementary charge $e$), $q_i$ denotes the ion valency, and $\rho_{b,i}$ is the reservoir density of each ionic species. We also define the potential of mean force (PMF)
\be
\Phi_i(\br;\kappa_v)\equiv-\ln\frac{\rho_i(\br)}{\rho_{b,i}}=\frac{q_i^2}2w(\br;\kappa_v)+q_i\phi_0
\label{PMF}
\ee
which is the change in the excess electro-chemical potential when an ion is brought from the bulk inside the pore at position $\br$ (in the following, all energies are rescaled by the thermal energy $k_BT=\beta^{-1}$). The quantity $w(r,\kappa_v)$ incorporates dielectric and electrostatic solvation forces (see Fig.~\ref{Cyl}) and thus depends in a complex manner on $\kappa_v$, and will be calculated in Section II. It is the $\kappa_v$ dependence of $w$ that couples $\phi_0$ to $\kappa_v$, without which $\phi_0$ is simply the Donnan potential. The variational scheme employed here, which handles ionic correlations and image forces at a non-linear level, was recently applied to charged slit pores~\cite{Nous}.  It was shown that the method i) is able to explore the regime between the WC  and SC limits for charged single and double interfaces and ii) goes beyond the self-consistent methods used in nanofiltration theories~\cite{yarosh,Nous}. The ability of the present method to interpolate between the  limiting WC and SC theories makes it a valuable tool for exploring the physics of charged biological systems (lipid membranes, water channels\ldots) where image forces play an extremely important role. Such a non-perturbative method is also essential for studying the possibility of an ionic \textit{liquid-vapor} (L-V) transition in a nanopore, our principal goal here ~\cite{prl}.
\textcolor{black}{In order to clarify the need for such a  method, we note that the for the weakly charged nanopores  investigated here the V-phase can to a very good approximation be assimilated to a counter-ion only phase in the SC limit, because the despite the low values of the SC parameter, $\Xi \sim 0.1$ the inequality (Eq.~4 of Jho et al, PRL 2008) delimiting the SC range of validity in a slab can, when extrapolated to a cylindrical nanopores of radius $\sim 1$~nm, still be satisfied thanks to the large value of the Gouy-Chapman length $\sim 10$~nm. Since the ionic L-phase is closer to the (salt) weak coupling limit, it is clear that a method is needed that spans the region between the two limiting laws.}

The variational equations obtained by Netz and Orland~\cite{netz} are equivalent to the closure equations established in the context of nanofiltration theories (see~\cite{loeb, Dresner} and \cite{yarosh} for a review). The closure equations were solved for spherical pores within a self-consistent approximation by Dresner~\cite{Dresner}, who observed a discontinuous phase transition from a high to a low ionic penetration state for decreasing reservoir concentration. Although the spherical geometry adopted by Dresner significantly simplifies the technical difficulties, it remains a toy-model since in reality,  ion-penetration is controlled by thin channels connecting the pore to the reservoir. The closure equations were later solved within the mid-point approximation by Yaroshchuk~\cite{yarosh} for a cylindrical pore. The mid-point approximation is equivalent to replacing the potential of mean force in the exponential of Eqs. (\ref{VarEq1}--\ref{VarEq2}) by its value in the middle of the pore, which leads to an underestimation of repulsive image forces and solvation deficit effects. As will be shown below, this approximation can overestimate the partition coefficients of ions in the pore by up to a factor of three.

This paper is organized as follows. We introduce in Section~II the general lines of the field-theoretic approach for charged dielectric bodies and derive the variational grand potential for a cylindrical pore. Section~III deals with ion penetration into a neutral pore of radius $a$.
A first-order ionic exclusion transition from an ionic-penetration state to an ionic-exclusion state is found for a specific range of parameter values. We draw the phase diagram in the $(a,\rho_b)$ space. We then compare our results with the self-consistent approach within the mid-point approximation~\cite{yarosh}. In order to compare the ionic penetration into slit~\cite{Nous} and cylindrical pores, we also compute the electrostatic potential and the variational free energy for two neutral concentric cylinders and study the limit of large cylinder radii with fixed separation. In Section~IV, we first investigate the effect of a non-zero fixed surface charge on ion partitioning and show that the transition survives for sufficiently weak surface charge density, but disappears above a critical value because of the additional cost in electrostatic energy of the pore surface charge, leading to an enhanced  ion penetration into the nanopore favorable to its screening. Finally, we develop our approach for the case where the surface charge density is regulated by the $p$H (surface charge regulation mechanism) which occurs experimentally for nanotubes in, e.g.,  PET membranes~\cite{lev} where weak acid groups are located at the pore surface.
The calculation of the Green's function for a cylindrical dielectric interface~\cite{JANCO} is presented in Appendix~A and for two concentric cylinders in Appendix~B.

\section{Model}
\begin{figure}[t]
\includegraphics[width=0.9\linewidth]{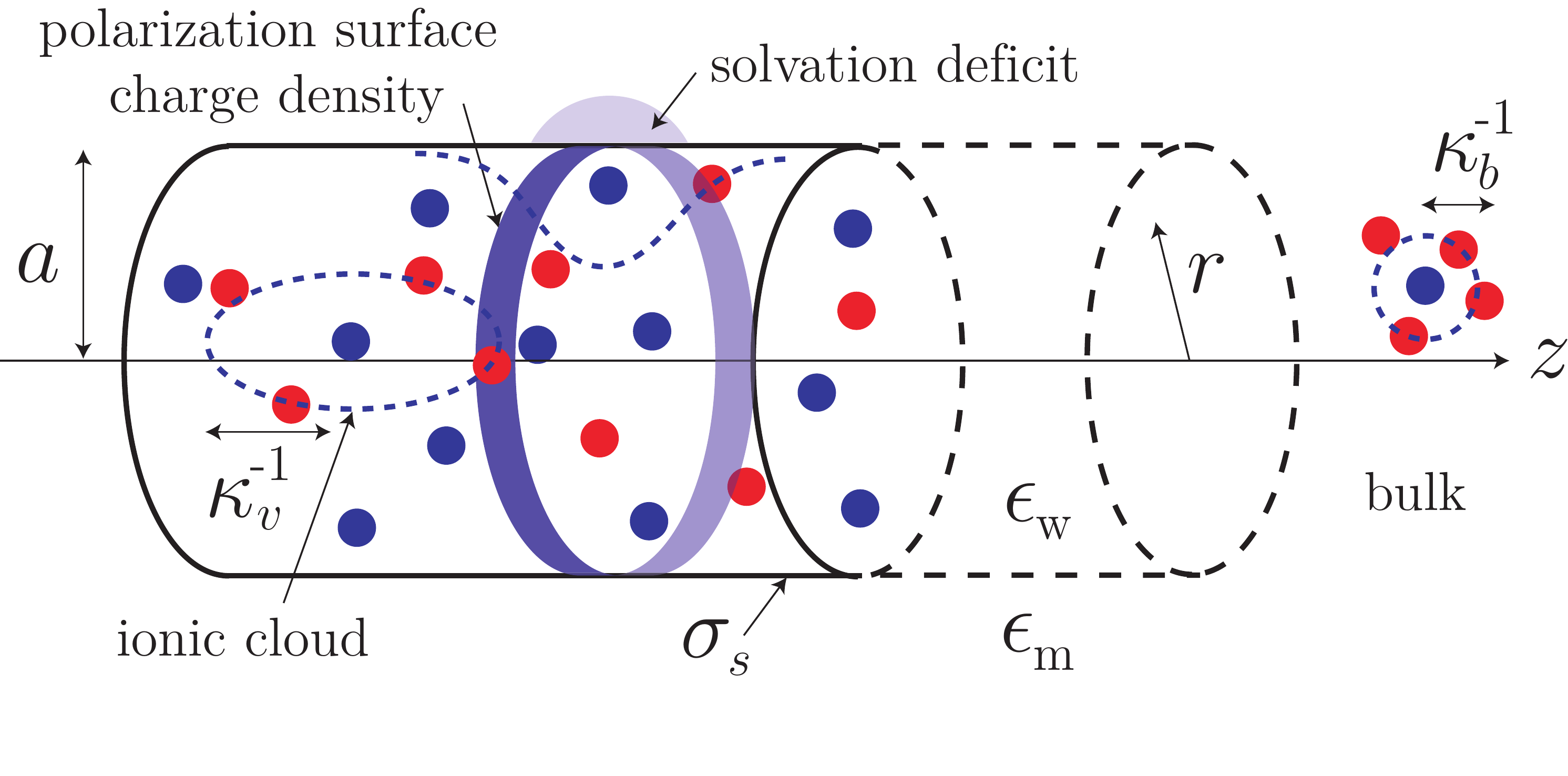}
\caption{Geometry for a cylindrical pore of infinite length and radius $a$. The cylindrical coordinates $(r,\theta,z)$ are defined in the figure. The ionic cloud is distorted due to the dielectric repulsion by the polarization surface charge density and the solvation deficit outside the pore.}
\label{Cyl}
\end{figure}
In this section, we compute the variational grand potential of a symmetric electrolyte of dielectric permittivity $\epsilon_{\rm w}$ (identical to the bulk value) confined in a cylinder of radius $a$ and length $L$. We consider in this article the limit of a sufficiently long cylinder ($L\gg a$) and neglect size effects. The cylindrical pore traversing a membrane is depicted in Fig.~\ref{Cyl}. The membrane of dielectric permittivity $\epsilon_{\rm m}$  is salt free (i.e. $\kappa=0$) and the electrolyte is in contact with an external particle reservoir at the end boundaries of the cylinder. Hence the fugacity of ions is $\lambda_i=e^{\mu_i}/\Lambda_i^3$, where $\Lambda_i=h/\sqrt{2\pi m_i k_BT}$ is the thermal De Broglie wavelength of an ion $i$ and $\mu_i$ its chemical potential, and it is fixed inside the pore according to the chemical equilibrium condition, $\lambda_i=\lambda_{i,b}$.

\begin{figure}[t]
\includegraphics[width=.9\linewidth]{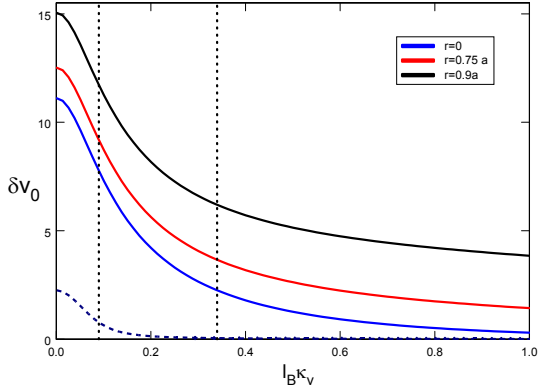}
\caption{Variation of $\delta v_0$ given in Eq.~(\ref{deltav0}) with $\kappa_v$ for three values of $r=0,0.75$, and $0.9\hspace{0.5mm}a$ (solid lines from bottom to top) and $\epsilon_{\rm m}=2$, $\epsilon_{\rm w}=78$, and $a=0.84$~nm. The dotted line corresponds to $\phi_0$ for $|\sigma_s|=5.4\times10^{-4}$~nm$^{-2}$ and $\rho_b=0.1812$ mol/L. The left and right reference lines mark the solutions of $\kappa_v$ just below and above the transition point, respectively. \label{Figdeltav0}}
\end{figure}

The grand-canonical partition function of $p$ species of interacting charges is
\be\label{ZG}
\mathcal Q=\prod_{i=1}^p\sum_{N_i= 0}^\infty\frac{\lambda_i^{N_i}}{N_i!}\int\prod_{j=1}^{N_i}\mathrm{d}\br_j\,e^{-(E_c-E_s)}
\ee
The electrostatic interaction in Eq.~(\ref{ZG}) is given by
\be
E_c=\frac12\int \mathrm{d}\br\mathrm{d}\br'\,\rho(\br)v_c(\br,\br')\rho(\br')
\ee
where the charge distribution, in units of the elementary charge $e$, is
\be
\rho(\br)=\sum_{i=1}^p\sum_{j=1}^{N_i}q_i\delta(\br-\br_j)+\rho_s(\br),
\ee
and $\rho_s(\br)=\sigma_s\delta(r-a)$ is the negative fixed charge distribution ($\sigma_s<0$) at the surface of the cylinder (1 $e$ nm$^{-2}=0.16$ C m$^{-2}$). The electrostatic potential $v_c(\br,\br')$ is solution of
\be\label{coulomb}
v_c^{-1}(\br,\br')=-\frac1{\beta e^2}\nabla\left[\epsilon(\br)\nabla\delta(\br-\br')\right]
\ee
where
\be
\epsilon(\br)=\epsilon_{\rm m} \Theta(r-a) + \epsilon_{\rm w} \Theta(a-r)
\ee
is the dielectric permittivity (where $\Theta(x)$ is the Heaviside distribution). Furthermore, the bulk self-energy of mobile ions that we substract from the total electrostatic energy in Eq.~(\ref{ZG}) is
\be
E_s=\frac{v^b_c(0)}{2}\sum_{i=1}^pN_iq_i^2.
\ee
The bare Coulomb potential in the bulk is $v^b_{\rm c}(\br)=\ell_B/r$, where the Bjerrum length is defined as $\ell_B=e^2/(4\pi\epsilon_{\rm w} k_B T)\simeq 0.7$~nm in water at $T=300$~K.
After performing a Hubbard-Stratonovitch transformation and summing over $N_i$ in Eq.~(\ref{ZG}), the grand-canonical partition function becomes $\mathcal{Q}=\int \mathcal{D}\phi\;e^{-H[\phi]}/Z_c$, where the field-theoretic Hamiltonian is
\be\label{ZEx}
H[\phi]=\int \mathrm{d}\br\left[\frac{\epsilon(\br)}{2\beta e^2}  [\nabla\phi(\br)]^2-i\sigma(\br)\phi(\br)-
\sum_i\tilde\lambda_i e^{i q_i \phi(\br)}\right]
\ee
where we have introduced the rescaled fugacities, $\tilde\lambda_i=\lambda_ie^{\frac{q_i^2}2 v_c^b(0)}$. The average electrostatic potential $\psi(\br)$ is related to the fluctuating one $\phi(\br)$ by $\psi(\br)=i\langle\phi(\br)\rangle$. The factor $Z_c$ in the partition function subtracts the non-screened van der Waals contribution. Since we are exclusively interested in the salt-dependent part of the grand potential, we keep $Z_c$ in the functional integral. The variational method consists in extremizing, with respect to the variational parameters, the first-order cumulant $\Omega_v=\Omega_0+\langle H-H_0\rangle_0$
where the expectation values $\langle\ldots\rangle_0$ are evaluated with the variational Gaussian Hamiltonian
\be\label{HVar}
H_0[\phi]=\frac12 \int_{\br,\br'}\left[\phi(\br)-i\phi_0(\br)\right]v^{-1}_0(\br,\br')\left[\phi(\br')-i\phi_0(\br')\right]
\ee
and $\Omega_0=-\frac12\mathrm{tr}\ln(v_0/v_c)$. The variational parameters are the Green's function $v_0(\br,\br')$ and the electrostatic potential $\phi_0(\br)$. In the following, we consider the restricted case of a constant variational ``Donnan'' potential $\phi_0$ and $v_0(\br,\br')$ a solution of the inhomogeneous variational Debye-H\"{u}ckel equation
\be\label{DHCY}
\left[-\nabla(\epsilon(\br)\nabla)+\epsilon(\br)\kappa^2(\br)\right]v_0(\br,\br')=\beta e^2\delta(\br-\br')
\ee
with a variational inverse screening length defined by
\be
\kappa(\br)=\kappa_v\,\Theta(a-r).
\ee
The restricted variational choice made here is both simple enough to lead to a tractable method and judicious enough to capture the essential physics for tight nanopores ($ a  <  2$~nm). After evaluating the functional integrals, the variational grand potential becomes
\begin{widetext}
\be
\frac{\Omega_v}{V_{\rm p}} = -\sum_i\lambda_i\lan \exp\left[\frac{q_i^2}2(\kappa_v\ell_B - \delta v_0(\br,\br;\kappa_v))-q_i\phi_0\right] \ran +\frac{\kappa_v^3}{24\pi} + \frac{\kappa_v^2}{8\pi \ell_B}\int_0^1 \mathrm{d}\xi\lan\delta v_0(\br,\br;\kappa_v\sqrt\xi)-\delta v_0(\br,\br;\kappa_v)\ran + \frac2a \sigma_s\phi_0
\label{omega}
\ee
\end{widetext}
where  $V_{\rm p}=\pi a^2L$ and $\delta v_0(\br,\br;\kappa_v)$ (see Fig.~\ref{Figdeltav0}) is the correction due to the presence of the nanopore to the variational Green function:
\be
v_0(\br,\br';\kappa_v)=\ell_B \frac{e^{-\kappa_v|\br-\br'|}}{|\br-\br'|} + \delta v_0(\br,\br';\kappa_v)
\ee
evaluated at $\br'=\br$ [see  Eq.~(\ref{GRCYin})], and is thus defined as
\be\label{deltav0}
\delta v_0(\br,\br;\kappa_v)=\frac{4\ell_B}{\pi}\int_0^\infty \mathrm{d}k\sideset{}{'}\sum_{m\geq0}F_m(k;\kappa_v)I_m^2(\varkappa |\br|)
\ee
where we note $\varkappa^2=k^2+\kappa_v^2$ and the prime on the summation sign means that the term $m=0$ is multiplied by 1/2. The function $F_m$ is a combination of modified Bessel functions $I_m$ and $K_m$ and is given in Eq.~(\ref{Fm}). Note that we have $F_m\to 0$ for $a\to\infty$ and thus $\delta v_0\to0$.

To find the physical meaning of the various contributions in Eq.~(\ref{omega}), it is interesting to rewrite it in two ways. The classical thermodynamic equality in presence of surface effects, $\Omega_v=-pV_{\rm p}+\gamma S_{\rm p}$ with $ S_{\rm p}=2\pi aL$, allows us to separate the volumic bulk contribution, the pressure, which is independent of $a$ (or $S_{\rm p}$)
\be\label{p}
p(\kappa_v)=\sum_i\lambda_i \exp\left(\frac{q_i^2\kappa_v\ell_B}2\right)-\frac{\kappa_v^3}{24\pi}
\ee
and a surface contribution term, $\gamma = (\Omega_v+pV_{\rm p})/S_{\rm p}$,
\begin{widetext}
\be
\gamma =\frac{a}2\sum_i  \lambda_i e^{q_i^2\kappa_v\ell_B/2} \lan 1 - e^{-q_i^2\delta v_0(\br,\br;\kappa_v)/2-q_i\phi_0}\ran + \frac{a\kappa_v^2}{16\pi \ell_B}\int_0^1 \mathrm{d}\xi\lan\delta v_0(\br,\br;\kappa_v\sqrt\xi)-\delta v_0(\br,\br;\kappa_v)\ran + \sigma_s\phi_0\label{gamma}
\ee
\end{widetext}
which is a function of nanopore characteristics ($\sigma_s$, $a$ and $\epsilon_m$ \textit{via} $\delta v_0$), and vanishes for $a\to\infty$.
By maximizing the variational pressure in Eq.~(\ref{p}), we find the inverse screening length in the bulk, $\kappa_b$, given by the following  implicit equation as a function of the fugacities $\lambda_i$:
\be\label{var_bulk}
\kappa_b^2=4\pi\ell_B\sum_i q_i^2 \lambda_i \exp\left(\frac{q_i^2\kappa_b\ell_B}2\right).
\ee
Note that Eq.~(\ref{var_bulk}) leads to instabilities for large values of $\lambda_i$ if hard-core repulsion is not included~\cite{note}. Moreover, Eq.~(\ref{var_bulk}) has no solution for large fugacities such that $q_i^2 \kappa_b(\lambda)\ell_B>4$. For low fugacities, Eq.~(\ref{var_bulk}) yields the Debye-H\"uckel Limiting Law (DHLL). Indeed, the bulk density is computed through
\be\label{rho_bulk}
\rho_{i,b}(\lambda) =\lambda_i\frac{\partial p}{\partial\lambda_i}= \lambda_i\, e^{\frac{q_i^2}2\kappa_b\ell_B}
\ee
which allows to rewrite Eq.~(\ref{var_bulk}) as
\be\label{DH}
\kappa_b^2=4\pi \ell_B \sum_i q_i^2\rho_{i,b}(\lambda).
\ee
In the following, we will keep $\rho_b$ in the equations by replacing $\lambda e^{\frac{q_i^2}2\kappa_b\ell_B}$ by $\rho_b$ in Eq.~(\ref{omega}). Note that to get $\rho_b$, we calculate it from $\lambda$, which is fixed in the grand canonical ensemble, using Eqs.~(\ref{var_bulk})-(\ref{rho_bulk}).
Solving Eq.~(\ref{rho_bulk}) for $\mu_i$ leads to the DHLL canonical ensemble result:
\be
\mu_i=\ln(\rho_{i,b}\Lambda_i^3)-\frac{q_i^2}2\kappa_b\ell_B.
\ee

The second way to interpret Eq.~(\ref{omega}) is by remarking that the first term on the rhs looks like the pressure of an ideal gas of ions ``dressed'' by their surrounding cloud, in a self-consistent or ``external'' field, $\frac{q_i^2}2(-\kappa_v\ell_B + \delta v_0(\br,\br;\kappa_v))+q_i\phi_0$. It favors high $\kappa_v$.
The second and third terms are the correlation contribution, or the charging energy needed to create dressed ions~\cite{mcquarrie}, modified by the nanopore. They come from ``attractive'' correlations between ions and their surrounding ionic cloud of opposite charge~\cite{mcquarrie}.  The second term is minus the DH pressure of a hypothetical bulk of inverse screening length $\kappa_v$ and the third one is a surface contribution due to the presence of the nanopore (dielectric exclusion and modified solvation effects). These two terms favor low $\kappa_v$ because the third term, $\Omega_d$ (Eq. A.15), like the second bulk one, increases with $\kappa_v$,  even if $\delta v_0(\kappa_v)$ is a decreasing function of $\kappa_v$ (see Fig~\ref{Figdeltav0}). If, on the one hand,  $\kappa_v$ is set arbitrarily to 0 (no screening, which corresponds to ``phantom ions''), one ends up with a simple barometric law, i.e. an ideal gas in the external potential, $\frac{q_i^2}2\delta v_0(\br,\br;0)+q_i\phi_0$, induced by the dielectric discontinuity at the pore surface. Within our restricted variational method this limit is the analog of the SC approximation (the ion-nanopore interactions dominate over the ion-ion correlations). If, on the other hand, we arbitrarily set $\kappa_v = \kappa_b \cosh  [\phi_0] \geq \kappa_b$ and neglect the influence of the dielectric discontinuity on  the Donnan potential $\phi_0$, we obtain the analog of the WC PB/DH theory. As will be shown below by an application of our variational method, these simplifying limits are not sufficient for the task at hand.

In summary, this system can be viewed as an ideal gas of dressed ions, whose accessible space is reduced  by dielectric repulsion. Both the Boltzmann weight, which takes into account the feedback correlation associated with the fact that ions forming the surrounding cloud are themselves dressed ions, and the repulsive dielectric self-energy contribution depend on the variational  screening parameter, $\kappa_v$.
The PMF, defined in Eq.~(\ref{PMF}), includes the potential $w(r)$ that incorporates the solvation and image-charge interactions
\bea\label{w}
w(r)&=&v_0(r,r)-v_{\rm c}^b(0)+\kappa_b(\lambda)\ell_B\\
&=& (\kappa_b-\kappa_v)\ell_B+\delta v_0(r,r;\kappa_v).\nonumber
\eea
where $\kappa_b$ is defined in Eq.~(\ref{var_bulk}).
The quantity $q_i^2w(r)/2$ is the difference between the excess chemical potential of ion $i$ located at distance $r$ in the nanopore and the excess chemical potential of the same ion in the bulk~\cite{Nous}. It is  a decreasing function of $\kappa_v$, since increasing $\kappa_v$ increases the electrostatic solvation gain of the hypothetic bulk (usual DH excess chemical potential  term in $-\kappa_v\ell_B$) and $\delta v_0(\kappa_v)$ decreases with $\kappa_v$ (very slowly for very low and high  $\kappa_v$ and abruptly at intermediate $\kappa_v \simeq 1/(2a)$ , see Fig.~\ref{Figdeltav0}), since the direct dielectric repulsion begins to be screened).

By minimizing Eq.~(\ref{omega}) with respect to $\kappa_v$, one exactly obtains Eq.~(\ref{VarEq1}) of the Introduction with $w(\br,\kappa_v)$ defined in Eq.~(\ref{w}).
Note that for a bulk electrolyte ($a\to\infty$) treated variationally, Eq.~(\ref{VarEq1}) yields the Debye-H\"uckel relation Eq.~(\ref{DH}) as long as the stability condition on fugacities $\lambda_i$, $q_i^2\kappa_b\ell_B<4$ ($\rho_b<3$ mol/L for monovalent ions) discussed in~\cite{curtis, Nous} is satisfied.
The drawback of working directly with Eq.~(\ref{VarEq1}) is that when it has three distinct solutions for $\kappa_v$, one cannot distinguish between stable, metastable and unstable solutions.
For this reason, in the following, we will analyze the pore-electrolyte model within the free energy minimization procedure and use Eq.~(\ref{VarEq1}) exclusively to show the importance of quantitative errors induced by a previous self-consistent approach~\cite{yarosh,Dresner}.

Finally, average ion concentrations in the pore are calculated using
\bea
\label{conc}
\lan\rho_i(r)\ran &=&-\lambda_i\frac{\partial (\Omega_v/V_{\rm p})}{\partial\lambda_i}\\
&=&\lambda_i\, e^{\frac{q_i^2}2\kappa_v\ell_B} \lan e^{-\frac{q_i^2}2\delta v_0(\br,\br;\kappa_v)-q_i\phi_0}\ran.
\eea
The ionic  partition coefficients, which are measurable quantities, for instance in ion conductivity experiments, are defined as
\be\label{PartCo}
k_i\equiv\frac{\left<\rho_i(r)\right>}{\rho_{i, b}}=\frac{2}{a^2}\int_0^{a} \mathrm{d}rr e^{-\frac{q_i^2}{2}w(r)-q_i\phi_0}.
\ee
In terms of the $k_i$, the first (entropic) term in $\Omega_v$  Eq.~(\ref{omega}), becomes $- \sum_i \rho_{i, b} k_i$.

In the rest of the paper, we will consider a symmetric electrolyte, i.e. $q_+=-q_-=q$. By differentiating the variational grand potential~(\ref{omega}) with respect to $\phi_0$, one simply obtains the electroneutrality condition Eq.~(\ref{VarEq2}) which for a cylindrical nanopore is simply
\be\label{electro}
\sigma_s=q\rho_ba\Gamma\sinh\left(q\phi_0\right)
\ee
where we have defined the coefficient
\be\label{Gamma}
\Gamma\equiv \lan e^{-\frac{q^2}{2}w(r)}\ran=2\int_0^1\mathrm{d}xx\hspace{0.5mm}\;e^{-\frac{q^2}{2}w(xa)}
\ee
which accounts for solvation and image corrections to mean-field theory (corresponding to $\Gamma=1$). By inverting Eq.~(\ref{electro}) we find for $\sigma_s < 0$
\be\label{solPhi}
q\phi_0=-\ln\left[\frac{q\Gamma/\alpha}{-1+\sqrt{1+(q\Gamma/\alpha)^2}}\right]
\ee
where we have defined $\alpha=|\sigma_s|/(\rho_ba)=X_m/(2\rho_b)$, $X_m=2|\sigma_s|/a$ being the volume charge density of the pore. The average potential $\phi_0$  increases in absolute value  as $1/q \ln [\alpha/(2q\Gamma)]$  for $\alpha/q\Gamma  \gg 1$. By injecting the solution  Eq.~(\ref{solPhi}) into the grand potential~(\ref{omega}), we are left with a single variational parameter $\kappa_v$ that will be varied in order to find the optimal solutions to the variational problem.

\section{Neutral pore}

In this section, we investigate the exclusion of ions from a neutral cylindrical pore. The electrolyte in the bulk reservoir is symmetric and composed of monovalent ions ($q=1$). For a symmetric electrolyte the vanishing surface charge, $\sigma_s=0$, imposes $\phi_0=0$ through Eq.~(\ref{electro}).
\begin{figure}[t]
(a)\includegraphics[width=.9\linewidth]{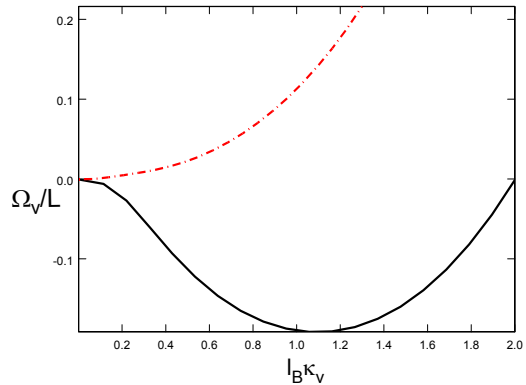}\vspace{-0.5cm}
(b)\includegraphics[width=.93\linewidth]{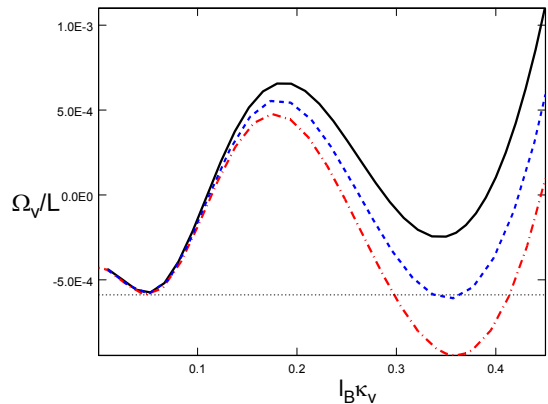}
\caption{Variational grand potential vs $\kappa_v$ for $\epsilon_{\rm m}=2$, $\epsilon_{\rm w}=78$ and $a=0.84$ nm. From bottom to top, bulk concentrations are (a) $\rho_b=0.77$ and 0.123~mol/L, and (b) $\rho_b=0.181$, 0.183, and 0.185~mol/L.\label{FreeI}}
\end{figure}
In Fig.~\ref{FreeI}a and~\ref{FreeI}b are plotted the variational grand potential $\Omega_v$ vs the adimensional variational inverse screening length $\ell_B\kappa_v$  for pore radius $a=0.84$~nm, $\epsilon_{\rm m}=2$ (the case for lipid membranes), and various bulk concentration $\rho_b$. Figure~\ref{FreeI}a shows that as one reduces the reservoir concentration from $\rho_b=0.77$~mol/L to $\rho_b=0.12$~mol/L, the minimum of the grand potential changes from $\ell_B\kappa^L_v\simeq1.1$  ($\lan\rho\ran=0.2524$~mol/L) to $\ell_B\kappa^V_v=0.037$ ($\left<\rho\right>=3\times10^{-4}$~mol/L). In other words, the pore evolves from an \textit{ionic-penetration liquid state} (L) to a quasi total \textit{ionic-exclusion vapor} one (V).

If one now slowly increases the bulk concentration from $\rho_b=0.12$~mol/L to, for instance, $\rho_b=0.181$~mol/L, one notices the apparition of a second minimum at moderate $\ell_B\kappa_v\simeq0.34$, shown in Fig.~\ref{FreeI}b, which corresponds to a significant ion concentration in the pore, $\rho_L=29$~mmol/L. One sees that this minimum is metastable and the stable solution is the one corresponding to the ionic-exclusion state with $\rho_b=0.44$~mmol/L. If we keep increasing the reservoir concentration up to $\rho^c_b=0.1832$~mol/L, the values of both minima become equal, which indicates a \textit{phase coexistence}: indeed, the equality of the two values of the grand potential, $\Omega_v(\kappa_v^V)=\Omega_v(\kappa_v^L)$, indicates mechanical equilibrium between both states. Finally, as $\rho_b$ is increased further to $\rho_b=0.185$~mol/L, the ionic-exclusion state becomes metastable and the pore becomes penetrable to ions.

This behaviour is the signature of a \textit{first-order phase transition} over a certain parameter range for cylindrical nanopores. Note that such a transition does not appear to take place in slit-like pores~\cite{Nous} although a continuous crossover from the presence to the absence of ions has already been observed for slit-like pores within a self-consistent calculation in Refs.~\cite{Dresner,yarosh} and within the variational approach in Ref.~\cite{hatlo,Nous} (see Figs.~7-9).
\begin{figure}[t]
\includegraphics[width=.96\linewidth]{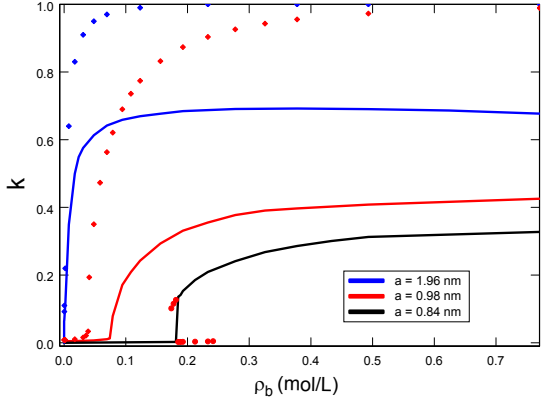}
\caption{Partition coefficients vs $\rho_b$ for $\epsilon_{\rm m}=2$, $\epsilon_{\rm w}=78$, and pore radii $a=1.96$, 0.98 and 0.84~nm (solid lines, from top to bottom). The (red) circles display the metastable solutions, thus defining the coexistence window. red and red diamonds correspond to the solution of the self-consistent  mid-point approximation Eq.~(\ref{MidPt}) for $a=1.96$, and 0.98~nm. \label{PartCoef2}}
\end{figure}

This discontinuous transition is characterized in Fig.~\ref{PartCoef2} where we display for $\epsilon_{\rm m}=2$ the partition coefficients, given by Eq.~(\ref{PartCo}), corresponding to the \textit{stable state} of the variational grand potential, as a function of the reservoir concentration $\rho_b$ for three different radii. At large pore size  $a=1.96$~nm, the discontinuous jump is absent and the transition to the ionic-exclusion state is a continuous crossover. At a critical pore radius  $a^*=0.987$~nm, a continuous phase transition occurs, where a single minimum exists for $\Omega_v$ which evolves towards lower ion concentrations in the pore with decreasing reservoir concentration in a fast but continuous way (data not shown).
For $a=0.98$~nm ($a<a^*$), the discontinuous jump at the coexistence bulk concentration exists, but with a very small jump for the partition coefficient.
\begin{figure}[t]
\includegraphics[width=.9\linewidth]{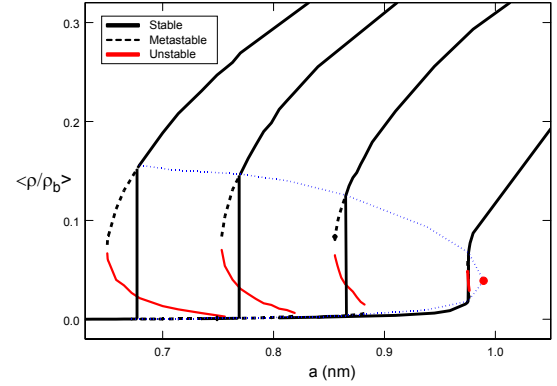}
\caption{Partition coefficient $k=\langle\rho\rangle/\rho_b$ inside a \textit{neutral} nanopore ($q=1$, $\epsilon_{\rm m}=2$, $\epsilon_{\rm w}=78$) vs the pore radius $a$ for, from left to right, $\rho_b=0.7,0.3,0.156,0.08$~mol/L (``isobars''). Dotted (grey/red) lines show metastable (unstable) branches, light grey/red lines (guide for the eye) are the ``boiling point'' curve (bottom) and the ``dew point'' curve (top) and the dot is the critical point, $\rho_b^*=0.075$~mol/L, $a^*=0.987$~nm.}
\label{Hysteresis}
\end{figure}

This is illustrated in Fig.~\ref{Hysteresis} where is shown the partition coefficient $k$ versus the pore size for various $\rho_b$. The metastable and unstable branches are shown in red and the critical point, where the transition becomes second order, is for $\rho_b^*=0.075$~mol/L, $a^*=0.987$~nm. One notices the resemblance with the liquid-vapor (L-V) phase transition in bulk fluids, showing the ``boiling point'' curve  and the ``dew point'' curve. Indeed, by interchanging the pore size, the partition coefficient of ions in the pore and the reservoir electrolyte density respectively with the temperature, the density and the pressure of a bulk liquid, one notices that the first-order ion exclusion transition in a nanopore becomes analogous to a bulk liquid-vapor (L-V) transition.
\begin{figure}[t]
\includegraphics[width=.9\linewidth]{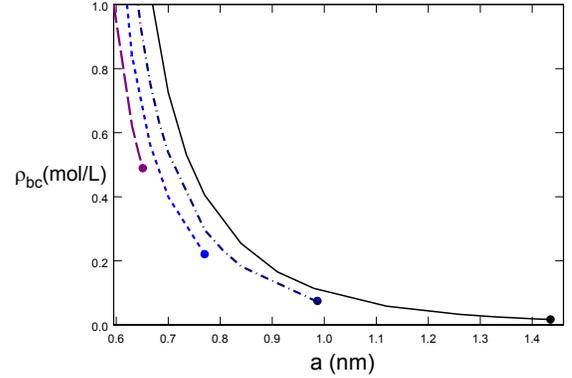}
\caption{Phase diagram characterizing the discontinuous phase transition for neutral pores ($\epsilon_w=78$ and $\epsilon_2=1,2,3,4$, from right to left). The critical line separates the ionic-penetration L state (area above the curve) from the ionic-exclusion V state (area below the curve) and ends at the critical point (dot) where the transition becomes continuous and then disappears.}
\label{PhaseDiagNt}
\end{figure}

To determine the parameter regime in which the discontinuous transition takes place, the phase diagram is shown in  Fig.~\ref{PhaseDiagNt}  for $\epsilon_{\rm w}=78$ and $\epsilon_{\rm m}=1$ to $\epsilon_{\rm m}=4$, where the lines correspond to the reservoir concentration at the coexistence point versus the pore radius, $\rho_b^c(a)$. Each curve corresponds to the coexistence line separating the ionic-penetration state, L, (the area above the curve) from ionic-exclusion state, V (below the curve). The lines end at the $\epsilon_m$ dependent critical point $(a^*,\rho_b^*)$, marked by a dot beyond which the transition disappears. One first notices that the parameter regime where the phase separation is observable is considerably reduced when increasing $\epsilon_{\rm m}$. The critical end-point evolves towards smaller pore sizes and higher reservoir densities. For $\epsilon_{\rm w}=78$ and $\epsilon_{\rm m}$ from 1 to 4 (the curves in Fig.~\ref{PhaseDiagNt}), the critical pore size is $a^*=1.44, 0.99, 0.77$ and 0.65~nm while the critical reservoir concentration is $\rho_b^*=17, 75, 221$ and 489~mmol/L, respectively. Higher values of $\epsilon_{\rm m}$ are not considered in this work since the coexistence line would correspond in this case to very high bulk densities and extremely small pore sizes, where hard-core effects, which are not yet incorporated in our model, become non-negligible.

\begin{figure}[t]
\includegraphics[width=.95\linewidth]{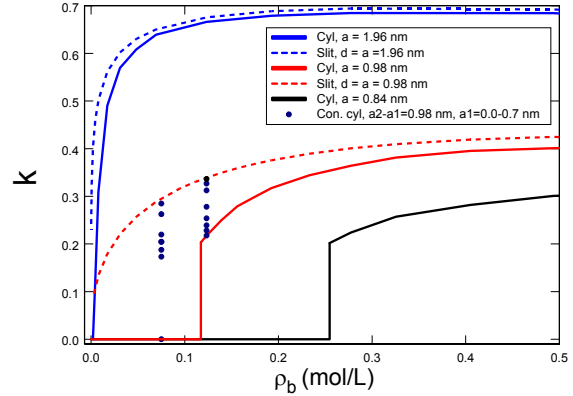}
\caption{Same figure as Fig.~\ref{PartCoef2} for $\epsilon_{\rm m}=1$. Solid lines correspond to a cylindrical pore and dotted lines to a slit pore whose width is $d=a$. Solid circles correspond to a the partition coefficient of an electrolyte confined between two concentric cylinders of a fixed width $\delta=a_2-a_1=a$ and show the crossover from a slit to cylindrical pore when $a_1$ varies from $\infty$ to 0. \label{PartCoef1}}
\end{figure}

It is clear from Fig.~\ref{PhaseDiagNt}  that the membrane dielectric permittivity $\epsilon_{\rm m}$, and thus dielectric repulsive forces, plays a central role in the mechanism of the phase transition. Moreover, it is known that the geometry influences the intensity of this dielectric repulsion. In order to investigate curvature effects on the ionic exclusion mechanism, we compare in Fig.~\ref{PartCoef1} the ion partition coefficients in cylindrical (continuous lines) and slit pores (dashed lines). It is seen that as long as one is far from the transition point, the ion density in a cylindrical pore of radius $a$ (diameter $2a$) is very close to that in a slit pore of thickness $d=a$ and the difference becomes smaller with increasing pore size $a$ and/or reservoir concentration $\rho_b$. In order to explain this equivalence, we computed within the variational approach the partition coefficient of ions confined between neutral concentric cylinders of radius $a_1$ and $a_2>a_1$ (technical details are given in Appendix B). The result is illustrated in Fig.~\ref{PartCoef1} for $\rho_b=0.075$~mol/L and $\rho_b=0.123$ mol/L by two sets of points computed at a fixed distance between the cylinders, i.e. $\delta=a_2-a_1=0.98$ nm. The inner radius $a_1$ is varied from top to bottom between 0.7 and 0~nm. At $\rho_b=0.123$~mol/L  and $a_1=0.7$~nm (the highest point), it is shown that the curvature of both cylinders is so weak that the ion concentration between the concentric cylinders is the same as that in the slit pore with $d=\delta$, as expected. By decreasing gradually $a_1$ from this value to 0, one naturally recovers the ionic partition function in a cylindrical pore. For $\rho_b=0.075$~mol/L, the interpolation between the slit pore and the cylindrical pore occurs in a similar way, except that since the pore radius $a=\delta$ corresponds to the exclusion state, the partition coefficient drops to nearly 0 between $a_1=0.119$~nm (the sixth point from the top) and $a_1=0.112$~nm (the seventh point from the top). Hence, we conclude that the equivalence in ionic penetration for cylindrical pores of radius $a$ and slit pores of thickness $d=a$ is simply the reminiscence of the concentric cylindrical case where for large values of the inner and outer radii (and the weak curvature of the dielectric interfaces), the system qualitatively behaves as a slit pore.

The existence of stronger dielectric forces in a cylindrical pore with respect to a slit one can thus be explained in terms of curvature. It is well-known that if the ion sees a dielectric interface with a positive curvature, i.e. if the interface is curved towards the ion, the experienced repulsion will be stronger than in the case where it would be placed face to a planar interface~\cite{Lue1}. On the contrary, if the ion is close to a negatively-curved interface, such as a protein or DNA, the dielectric exclusion will be weaker compared to a planar interface. This is the reason why the cavity correction factor of proteins measured in experiments is below one~\cite{DiExp}. Therefore, the positive curvature of the pore is the essential ingredient for the existence of a discontinuous phase transition.

At the first-order variational level [see Eq.~(\ref{omega})], ion penetration is essentially driven by two opposing mechanisms which contribute to the surface term $\gamma$, given in Eq.~(\ref{gamma}). On the one hand, due to \textit{``attractive'' correlations} between ions and their oppositely charged surrounding cloud, both the equivalent bulk term, $\kappa_v^3/(24 \pi)$, and the $\kappa_v^2\int \mathrm{d}\xi\lan \delta v_0(\br,\br,\kappa_v\sqrt\xi)-\delta v_0(\br,\br,\kappa_v)\ran$ term   vanish for $\kappa_v=0$ and increase with  $\kappa_v$ (the second term  being sensitive to  variations of $\kappa_v$, because increasing $\kappa_v$ leads to the screening of $\delta v_0$, see Fig.~\ref{Figdeltav0}). This cost in surface contribution is thus associated with the deformation of the ionic cloud of each ion due to dielectric repulsion and favors low  $\kappa_v$.

On the other hand, for strong dielectric exclusion and weak surface charge, the presence of the pore walls decreases the volume accessible to ions compared to an hypothetic bulk, and therefore the first (depletion) term in Eq.~(\ref{gamma}) contributes positively to the surface tension and decreases in magnitude with increasing $\kappa_v$ . When $\kappa_v$ increases, the dielectric repulsion becomes screened and the number of ions penetrating the nanopore increases (due to the Boltzmann law), which decreases this surface cost. Indeed, the Gibbs adsorption equation at constant temperature is
\be
\mathrm{d}\gamma = -\left(\lan\rho\ran-\rho_b\right) \mathrm{d}\mu.
\ee
By increasing $\kappa_v$, Eq.~(\ref{w}) leads to $\mathrm{d}\mu<0$, and since there is desorption in the nanopore ($\lan\rho\ran<\rho_b$), $\mathrm{d}\gamma<0$: this is favorable to the system and thus favors high  $\kappa_v$.

Note that this liquid-vapor phase transition exists for bulk electrolytes but at very low temperature ($T\approx100$~K) for usual mineral salts~\cite{stell,singh,fisher,hynn}. The critical temperature $T^*$ is thus shifted towards room temperature in pores of nanometer radius due to surface effects. Indeed, essentially due to dielectric repulsion, the low dielectric surface favors the vapor phase and thus shifts the first-order transition. We thus call this phenomenon \textit{capillary evaporation}, just as for  nano-capillary confined  water, which undergoes condensation for hydrophilic surfaces~\cite{evans,evans_review,gubbins,peterson} and evaporation for hydrophobic ones~\cite{roth,beckstein}.

The possibility for the existence of a first-order phase transition for electrolytes confined in \textit{spherical} pores was first pointed-out by Dresner~\cite{Dresner}. He used an approximate Green's function, which is exact at the center of a spherical pore, and a self-consistent approach to show that the first-order phase transition survives as long as $\epsilon_{\rm m}\ll\epsilon_{\rm w}$. However, because of the numerous approximations introduced in his work, Dresner~\cite{Dresner} recognized that the quantitative predictions of this approach were not reliable. Yaroschuk~\cite{yarosh} later performed a similar self-consistent calculation using the correct Green's function at the center of a \textit{cylindrical} pore and showed that in this case a  discontinuous phase transition to an ionic-exclusion state also coccurs. He also argued, however,  that this phase transition is unphysical because it comes from the use of the DH equation, which ignores non-linear effects. By introducing a non-linear DH equation (in analogy with the non-linear PB equation), he argued that the first-order phase transition should disappear.
However, this PB-like generalization of the DH equation is incorrect as already argued by Onsager in the 1930s~\cite{onsager}. Hence, we argue that, within the confined electrolyte model presented in this work, the existence of a first-order phase transition is physically sound.

We finally analyze Yaroschuk's self-consistent (SC) approach. It consists in finding self-consistently the screening parameter $\kappa$ by defining $\kappa^2$ as the average value of $\kappa_b^2\;e^{-\Phi(\br,\kappa)}$ in the pore~\cite{yarosh}. To pursue in a cylindrical geometry, he replaced the PMF within the integral by its value on the pore-axis, which in turn causes an underestimation of repulsive dielectric forces that become significantly stronger close to the pore wall. He obtained what we call the  ``mid-point approximation'' for $\kappa$:
\be\label{MidPt}
\kappa^2=\kappa_b^2\;e^{-\frac{q^2}2 w(0,\kappa)}
\ee
where $\kappa_b$ is defined in Eq.~(\ref{DH}). The partition coefficients in a neutral pore obtained from the numerical solution of Eq.~(\ref{MidPt}) are compared in Fig.~\ref{PartCoef2} with the prediction of our variational approach for $a=1.96$ and 0.98~nm. It is clearly seen that the mid-point approximation overestimates ionic penetration into the pore, which is due to the underestimation of dielectric forces, as stressed above.
This point was also noticed for slit-pores in our previous work~\cite{Nous}. As one sees in Fig.~\ref{PartCoef2}, the error induced by the mid-point approach increases with decreasing pore size, which originates from the amplification of image forces when one decreases the pore radius. One notices that for $a=0.98$ nm  the partition coefficients as well as the position of the transition point predicted by the mid-point approximation can deviate from the predictions of our variational method by 200--300 $\%$.

\section{Charged pore}

\subsection{Fixed surface charge}

In this part, we investigate ion penetration into a cylindrical nanopore of non-zero negative fixed surface charge density, $\sigma_s<0$. We first note that the electroneutrality condition Eq.~(\ref{electro}) can be written, using Eq.~(\ref{PartCo}), as
\be\label{DifPart}
k_+-k_-=\frac{2|\sigma_s|}{q\rho_ba}=\frac{X_m}{q\rho_b}.
\ee
We introduce the \textit{good coion exclusion} (GCE) limit which corresponds to the counterion-only case in the nanopore. In the case of slit pores, it was shown that the \textit{charge repulsion} GCE limit is reached for small pore sizes, strong surface charge, and low bulk concentrations even without any dielectric effects~\cite{Nous} ($\Gamma\approx1$, $2|\sigma_s|/(q\rho_ba)\gg1$) and associated with the electric repulsion of coions. In the case of cylindrical nanopores, a second \textit{dielectric repulsion} GCE limit exists for low $|\sigma_s|$ as soon as $\Gamma\ll2|\sigma_s|/(q\rho_ba)$. The partition coefficients of ions in the cylindrical pore become
\be\label{GCE}
k_+\simeq\frac{2|\sigma_s|}{q\rho_ba}\quad\mathrm{and}\quad k_-\simeq\frac{qa\rho_b}{2|\sigma_s|}\Gamma^2\ll k_+.
\ee
The first equality shows that in the  GCE regime, the average counterion concentration in the pore, $\langle\rho_+\rangle=k_+\rho_b$, is independent of $\rho_b$ and depends only on the pore radius $a$ and the surface charge $\sigma_s$. Hence, we are in a regime where dielectric repulsion plays only an indirect role (by excluding  coions) in determining the average nanopore counterion concentration, which is solely determined by the global electroneutrality in the pore.
\begin{figure}[t]
\includegraphics[width=.9\linewidth]{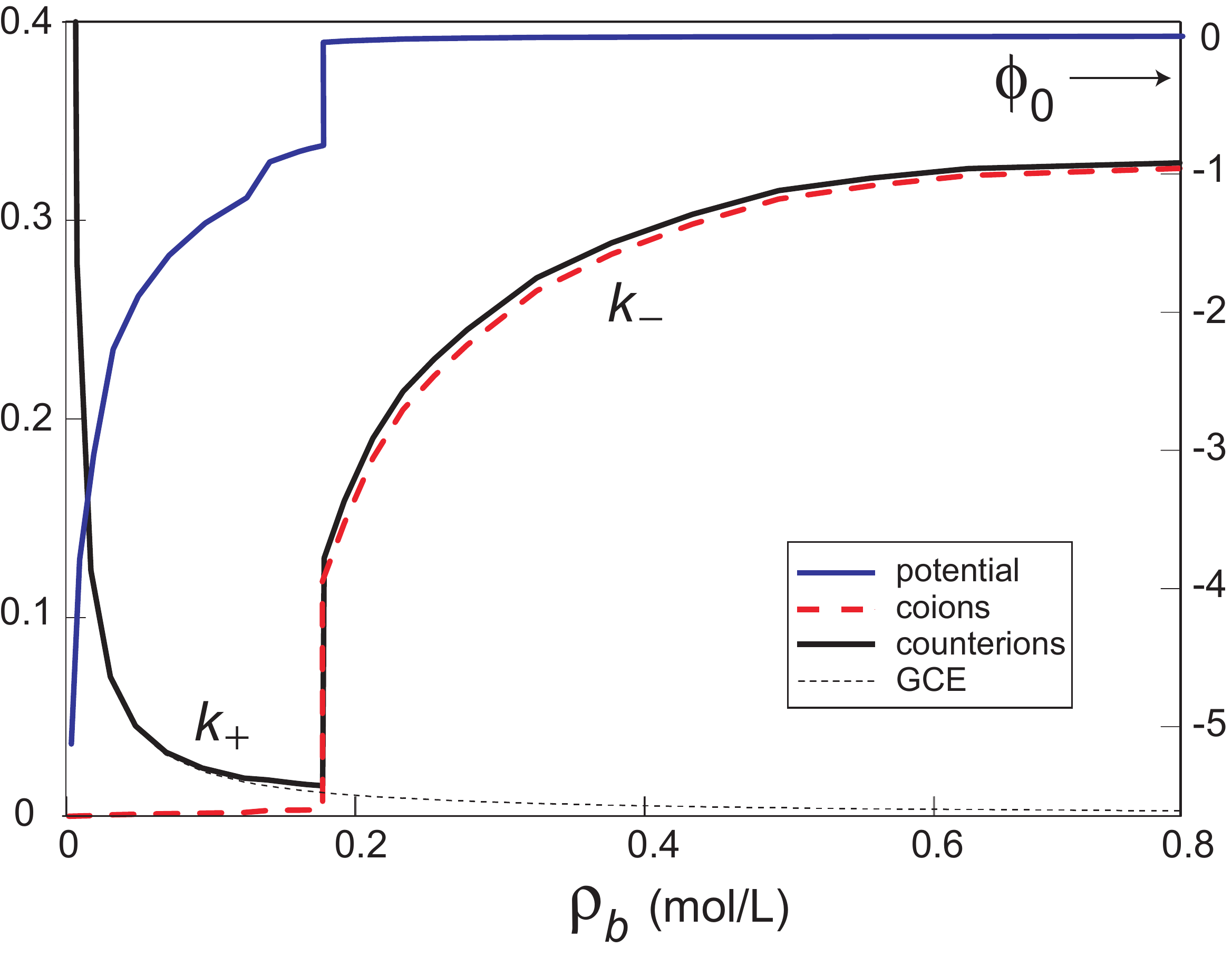}
\caption{Partition coefficients of coions and counterions versus the reservoir concentration (above) and the effective Donnan potential (below) for $\epsilon_{\rm m}=2$, $\epsilon_{\rm w}=78$, $a=0.84$ nm and $|\sigma_s|=5.4\times10^{-4}$~nm$^{-2}$. The dotted line in the top plot denotes the GCE regime Eq.~(\ref{GCE}).}
\label{CoefPartCH1}
\end{figure}

We plot in Fig.~\ref{CoefPartCH1} the partition coefficient of ions $k_\pm$ and the variational Donnan potential $\phi_0$ against the reservoir concentration $\rho_b$ for $\epsilon_{\rm m}=2$, $a=0.84$~nm and a weak surface charge, $|\sigma_s|=5.4\times10^{-4}$~nm$^{-2}$. One observes that at high bulk concentrations, $k_+$ and $k_-$ decrease when $\rho_b$ decreases, as in the neutral case [and indeed $\phi_0\approx0$ since $q\Gamma\gg \alpha$ in Eq.~(\ref{solPhi}), the slight difference $k_+-k_-$ being given by Eq.~(\ref{DifPart})] down to a characteristic value $\rho_b^c$ where a discontinuous jump towards a \textit{weak ionic-penetration} state takes place. It is interesting to note that the Donnan potential $\phi_0$ also exhibits  a jump at this point. A first conclusion is the existence of a discontinuous transition for charged pores at low $|\sigma_s|$, with the coexistence value $\rho_b^c(\sigma_s)\lessapprox\rho_b^c(\sigma_s=0)$. If one further decreases the reservoir concentration, $k_-$ evolves towards vanishingly small values while $k_+$ abruptly changes by rapidly increasing. We thus reach the dielectric repulsion GCE regime, Eq.~(\ref{GCE}), denoted by the dotted curve. Although the concentration of ions is low, it is not zero (in order to fulfill electroneutrality) and the vapor phase is thus now a  \textit{weak ionic-penetration} phase. In this regime, $|\phi_0|$ increases abruptly (as $1/q \ln [\alpha/(2q\Gamma)]$) because the dielectric exclusion dominates and thus $q\Gamma\ll \alpha$ in Eq.~(\ref{solPhi}).

\begin{figure}[t]
(a)\includegraphics[width=.9\linewidth]{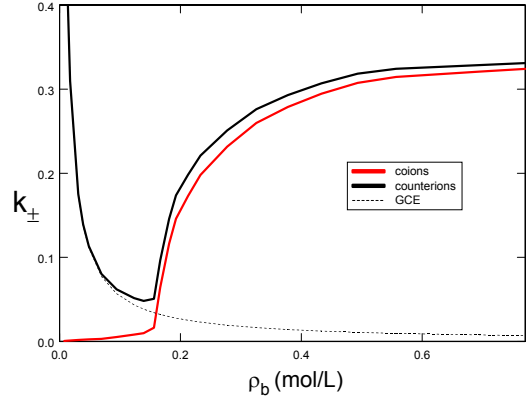}
(b)\includegraphics[width=.92\linewidth]{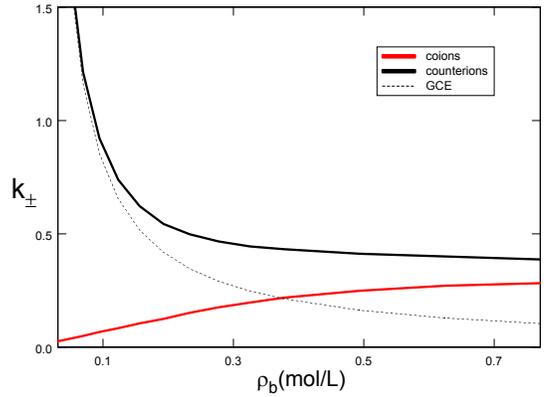}
\caption{Partition coefficients of coions and counterions versus the reservoir concentration  for $\epsilon_{\rm m}=2$, $\epsilon_{\rm w}=78$, $a=0.84$~nm and $|\sigma_s|=1.35\times10^{-3}$~nm$^{-2}$ (above) and $|\sigma_s|=2\times10^{-2}$~nm$^{-2}$ (below). The dotted line in the top plot denotes the GCE regime Eq.~(\ref{GCE}).}
\label{CoefPartCH2}
\end{figure}

Figure~\ref{CoefPartCH2}a shows the partition coefficients for a slightly stronger surface charge density $|\sigma_s|=1.35\times10^{-3}$~nm$^{-2}$. In this case, the discontinuous jump disappears and the transition becomes continuous. Hence a large enough fixed surface charge density destroys the first-order phase transition. Figure~\ref{CoefPartCH2}b shows the partition coefficients for a higher surface charge $|\sigma_s|=2\times10^{-2}$~nm$^{-2}$. In this case, not only the discontinuous phase transition, but also the turning point characterizing the abrupt change in the behaviour of $k_+$ disappears from the range of displayed $\rho_b$. In other words, above a characteristic surface charge and at low concentrations, the dielectric repulsion GCE regime sets in and the counterion penetration into the pore is solely determined by the global electroneutrality Eq.~(\ref{GCE}).

\begin{figure}[t]
\includegraphics[width=.9\linewidth]{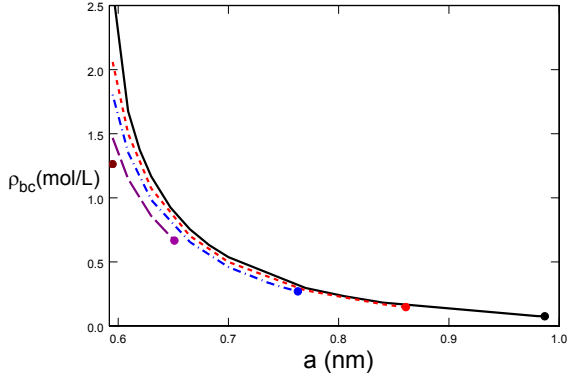}
\caption{Phase diagram characterizing the discontinuous phase transition in charged pores for several values of the fixed surface charge. The critical line separates the ion-penetration state (area above the curve) from the ion-exclusion state (area below the curve) and ends at the critical point (dot). $\epsilon_{\rm m}=2$ and $\epsilon_{\rm w}=78$. From top to bottom, the lines correspond to $|\sigma_s|=0,1,2,4,6 \times10^{-3}$~nm$^{-2}$.}
\label{PhaseDiagCh}
\end{figure}

In the previous section, we emphasized that the first-order nature of the transition is associated with the variations of $\delta v_0$ with $\kappa_v$, which influences both the nanopore modified DH correlations (which favor the low density V phase) and the entropic term (which favors the high density L phase). The absence of a first-order phase transition when increasing $\sigma_s$ beyond a characteristic value is due to the increase of the surface contribution  associated with the positive electrostatic energy of the pore surface charge  (the last  term in Eqs.~(\ref{omega}, \ref{gamma})), which favors the higher ion concentration in the nanopore necessary for screening out the pore surface charge. The complete characterization of the phase transition is illustrated in Fig.~\ref{PhaseDiagCh} where is shown the phase diagram for several values of $\sigma_s$. The main effect of the surface charge is to reduce the coexistence line and to shift the critical point towards smaller pore sizes and higher reservoir densities. Comparison of Figs.~\ref{PhaseDiagCh} and~\ref{PhaseDiagNt} clearly shows that the increase of the surface charge plays qualitatively the same role as an increase of the membrane dielectric permittivity. It is important to note that the smearing of the discontinuous phase transition takes place over a very narrow surface charge range: within the parameter regime considered in Fig.~\ref{PhaseDiagCh}, the transition completely disappears for $|\sigma_s|>6.0\times10^{-3}$~nm$^{-2}$.

It is well known that nature uses the very same pore surface chargemechanism to deal with strong image forces in water-filled ion channels~\cite{Hille}. As it was illustrated in the Section~III, a neutral channel is impermeable to ions at low reservoir concentrations. In potassium channels, negatively charged carbonyl oxygens located at the pore surface~\cite{Kamen,Kameneev} reduce the potential barrier induced by image interactions and make the channel cation-selective.

\subsection{Charge regulation mechanism}

In the previous part, we considered a boundary surface characterized by a fixed and uniform surface charge. However, in experiments on the conductivity of nanopores in PET membranes~\cite{lev,pasternak}, the surface charge density is unknown and cannot be measured easily.  If one wishes to make quantitative comparisons with these experiments, \textcolor{black}{one should consider the fact that, in these experiments where the pore surface carries carboxylic acid groups, $\sigma_s$ increases with $p$H~\cite{hijnen}. The $p$H in the reservoir, defined as $p\mbox{H}=-\log_{10}\rho_b\left(\mathrm{H}_3\mathrm{O}^+\right)$, is thus used as a control parameter.}

\textcolor{black}{
In this section, we introduce a charge regulation mechanism where, in the presence of water, chemical groups located at the surface may become chemically active, leading to proton release from the acid groups and thus to a variable surface charge. As a first hint of the full problem, we use two approximations: i) we consider exclusively the case of trace hydronium ions, $\mathrm{H}_3\mathrm{O}^+$, (spectator ions) with a bulk concentration ($2<p\mathrm{H}<12$) significantly lower than that of salt ions; ii) the effective surface charge density, which becomes a function of the $p$H, is computed at the an effective mean-field level ~\cite{PodChrReg}, i.e. we  consider that image forces, etc. for the hydronium ions in the pore and thermal fluctuations of the electric potential are implicitly accounted for via en effective equilibrium constant.
}

The charge dissociation mechanism at the surface is described by the following chemical equilibrium:
\be\label{ChemI}
\mathrm{[surface]-COOH}+\mathrm{H}_2\mathrm{O}\leftrightarrows \mathrm{[surface]-COO}^-+\mathrm{H}_3\mathrm{O}^+
\ee
characterized by the equilibrium constant $K_a$. \textcolor{black}{Note the this effective $K_a$ implicitly takes into account not only  image forces but also all other non-electrostatic interactions at the pore surface.}
\textcolor{black}{The source term [last term of Eq.~(\ref{omega})] is thus modified. In a simple Langmuir two-state model~\cite{ninham,PodChrReg}, each carboxylic group at the cylindric surface is a site, with surface density $\sigma_0$, which can be empty, i.e. \textit{dissociated}, with an energy per site $-\phi_0$, or occupied, i.e. \textit{associated with an hydronium ion}, with an energy $E_a$ which takes into account the chemical bounding and all other surface interactions and which, in a first approximation, is taken to be independent of $\kappa_v$. The grand potential for this adsorbed two-dimensional hydronium gas is then
\be
\mathcal{Q}_{2D}=e^{\phi_0}+\lambda^+e^{E_a}
\ee
where $\lambda^+=\lambda_b^+=\ln\rho_{b,\mathrm{H}_3\mathrm{O}^+}e^{-\kappa_b\ell_b/2}$ is the fugacity of hydronium ions and $E_a$ is the energy gained in the associated state. The equilibrium constant of Eq.~(\ref{ChemI}) is thus defined as $K_a\equiv\frac{\gamma_{\mathrm{H}_3\mathrm{O}^+} \gamma_{\mathrm{COO}^-}}{\gamma_{\mathrm{COOH}}}=e^{-E_a+\kappa_b\ell_B}$, where $\gamma_i$ are activities coefficients of the relevant species.}

\textcolor{black}{Instead of $\sigma_s\phi_0$ in Eq.~(\ref{omega}), the source term becomes now~\cite{PodChrReg}
\bea
\frac{\Omega_{2D}}{S} &=& -\sigma_0\ln\mathcal{Q}_{2D} \label{ChrReg}\\
&=&-\sigma_0\phi_0-\sigma_0\ln\left[1+10^{p\mathrm{Ka}-p\mathrm{H}}\;e^{-\phi_0}\right]\nonumber
\eea
where $p\mathrm{Ka}=-\log_{10}K_a$. By minimizing the full grand potential with respect to $\phi_0$, one obtains the electroneutrality condition Eq.~(\ref{electro}) with the effective surface charge density~\cite{Manciu}
\be\label{SigmaH}
\sigma_s=-\sigma_0-\frac{\lambda^+}S\frac{\partial \Omega_{2D}}{\partial \lambda^+}=\frac{-\sigma_0}{1+10^{p\mathrm{Ka}-p\mathrm{H}}\;e^{-\phi_0}}.
\ee
}

\begin{figure}[t]
(a)\includegraphics[width=.9\linewidth]{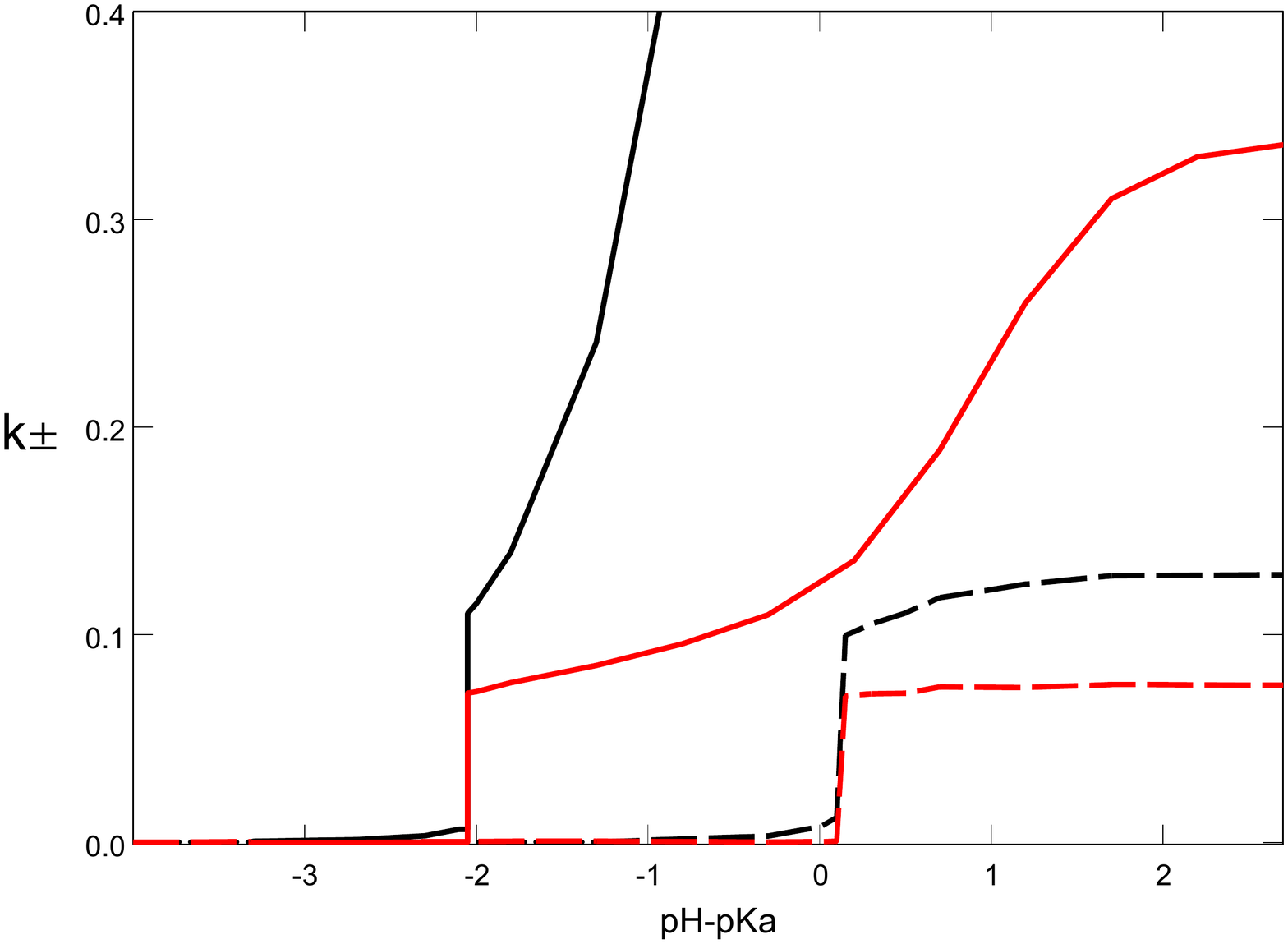}
(b)\includegraphics[width=.93\linewidth]{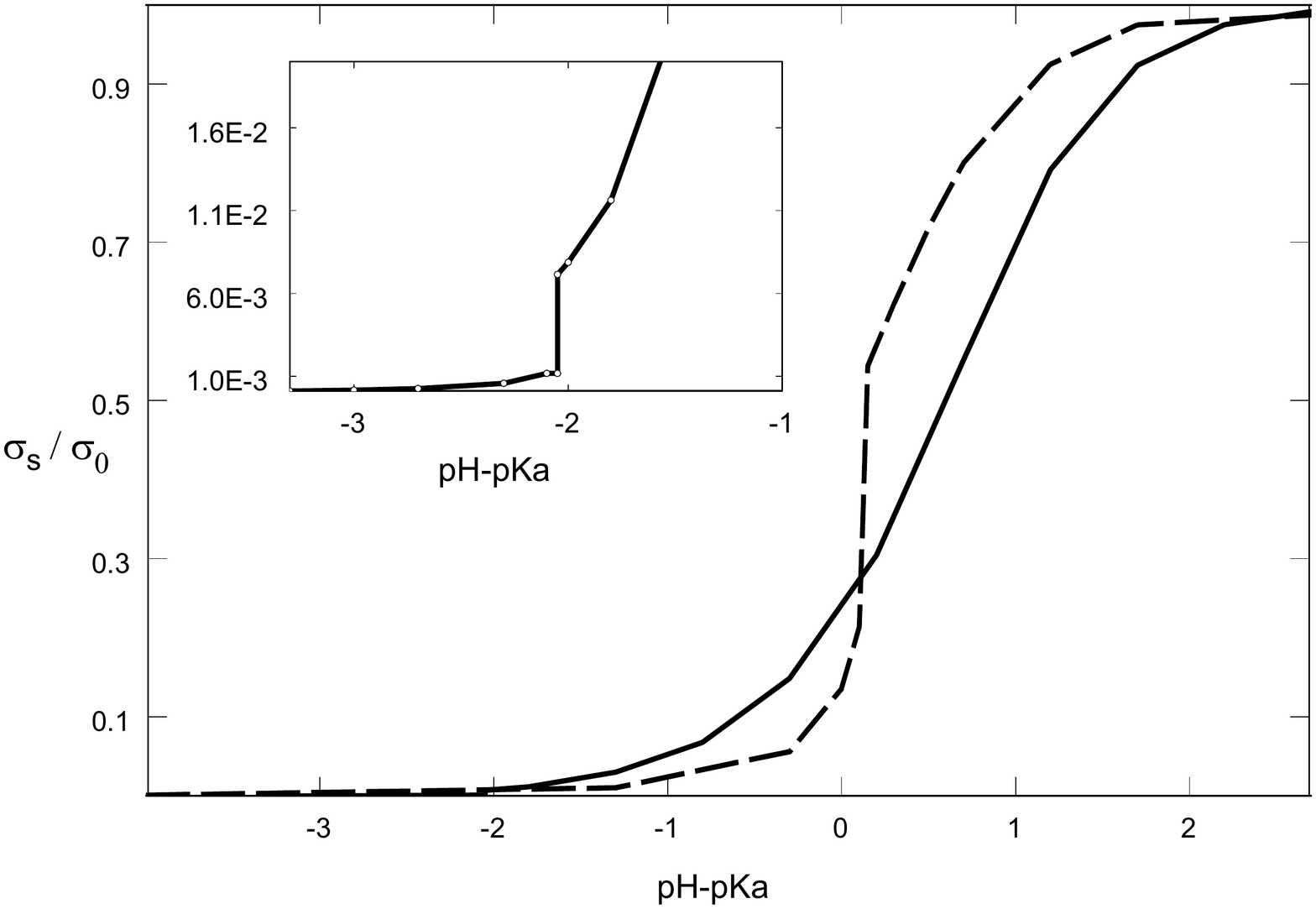}
\caption{(a) Partition coefficients of cations (black lines) and anions (red lines) and (b) the reduced surface charge versus $p$H-$p$Ka for $\epsilon_m=2$, $\epsilon_{\rm w}=78$, $\rho_b=1$~mol/L and $a=0.617$~nm. Solid lines correspond to $|\sigma_0|=1$~nm$^{-2}$ and dashed lines to $\sigma_0=10^{-2}$~nm$^{-2}$. The inset in the bottom plot illustrates the jump in the net surface charge for $\sigma_0=1$~nm$^{-2}$.}
\label{ChrReg}
\end{figure}
Figure~\ref{ChrReg} illustrates for $\sigma_0=1$ and 10~nm$^{-2}$ the ionic partition coefficients (Fig.~\ref{ChrReg}a) and the surface charge (Fig.~\ref{ChrReg}b) versus the acidity of the solvent, $p\mathrm{H}-p\mathrm{Ka}$. The low $p$H regime corresponds to the neutral pore limit, where hydronium ions of high bulk concentration almost totally neutralize the acid groups of the surface and the system is in the weak ionic penetration state. By increasing $p$H (or decreasing the hydronium concentration in the bulk), the partition coefficient slowly increases until a characteristic value, $p$H$^C$, beyond which one crosses the critical line and due to the high surface charge attraction, the system evolves into the ionic penetration state. This transition is illustrated in Fig.~\ref{ChrReg}a by the sudden jump of the coion and counterion partition coefficients, $k_\pm$. The most remarkable prediction of our charge regulation model is that  the surface charge density displays a sharp discontinuous \textcolor{black}{increase at the transition point. This behavior is obviously associated with the jump of the electrostatic potential $\phi_0$ (see Fig.~\ref{CoefPartCH1}) at the transition from the V- to the L-phase: salt ions induce the screening of the electrostatic potential in the pore ¤$\phi_0\simeq0$, which leads to a release of the spectator hydronium ions to the bulk and thus a sudden increase of the surface charge.}

The variations of $k_\pm$ with $p$H discussed above can explain the fluctuations found in experiments carried out on the conductivity of nanopores in PET membranes~\cite{lev,pasternak}, where a rapid switching between a high conductivity (HC) and low conductivity (LC) regime was observed. We argue that these fluctuations may correspond to the system being close to or at phase coexistence  between both regimes, where the HC state would correspond to the liquid  strong ionic penetration  state and the LC state to the weak ionic penetration vapor state predicted within our nanopore model (we present a  more detailed theoretical investigation of electric current fluctuations within our nanopore model  elsewhere~\cite{prl}).  Furthermore,   HC/LC switching is  observed experimentally only within a  narrow $p$H window and surface charge fluctuations are observed to be strongly correlated with conductivity ones ~\cite{lev}, both of which can be accounted for using our model, see Fig.~\ref{ChrReg} b. and the inset of Fig. 3b of \cite{prl}. Finally, it would be interesting to check experimentally our  theoretical prediction for the ratio of the surface charge values, $\sigma_s^{L}/\sigma_s^{V}\simeq 0.51$, at coexistence.

\section{Conclusion}

In this article, we developed a variational approach to investigate ion penetration into \textit{cylindrical} nanopores. The variational scheme has already been developed in Ref.~\cite{Nous} and is based on a generalized Onsager-Samaras approximation which assumes a \textit{constant effective variational screening length} and a \textit{uniform Donnan potential} enforcing electroneutrality in the pore. These two physically sound simplifications to the full variational problem allow us to handle the more complicated case of ions confined in a charged dielectric nanopore.

In the first part of the work, we considered ionic exclusion from a neutral dielectric nanopore. It is shown that, when the electrolyte concentration in the reservoir or the pore size is decreased down to $\rho_b\simeq 0.1$~mol/L or  $a\simeq 1$~nm (see Fig.~\ref{PhaseDiagNt}), a first-order phase transition from an ionic-penetration state to an ionic-exclusion one emerges. The underlying physical mechanism this transition  is a competition between Gibbs adsorption which decreases the surface tension by favoring strong ion penetration  and  DH-type attractive correlations strengthened by the presence of the dielectric nanopore wall, which favor low ion penetration.

By comparing our predictions to the ones obtained using a self-consistent mid-point approach~\cite{yarosh}, it is shown that the latter can over-estimate ion concentration by a factor of 2 to 3. We also propose a continuous description from the slit geometry (thickness $d$), where no transition but rather a continuous crossover appears~\cite{Nous}, to the cylindrical one (diameter $2d$), by investigating  the case of concentric cylinders (separated by a distance $d$), thus highlighting the role of dielectric surface curvature.

In the second part, we investigated the effect of a uniform surface charge density on this ionic-exclusion phase transition.
When varying bulk concentration, the counterion penetration into the cylindrical nanopore, induced by the charged surface in order to fulfill  electroneutrality, exhibits a non-monotonic behaviour characterized by two regimes: i) a weak ionic-penetration (or dielectric repulsion GCE) regime, for low reservoir concentrations or small pore sizes, where the decrease of the coion partition coefficient with decreasing bulk salt concentration  is even stronger than for a neutral pore, due to dielectric exclusion combined with electrostatic repulsion, and the counterion concentration remains low and is independent of $\rho_b$. ii) A second neutral-pore like ion penetration regime, reached above a characteristic reservoir concentration, where repulsive image forces  are lower and the electrostatic potential is almost zero. The counter-ion partition coefficient $k_+$ increases with the reservoir concentration.
For very small surface charge densities ($<2\times10^{-3}$~nm$^{-2}$), the change between both regimes corresponds to a first-order phase transition.
Increasing the surface charge density shrinks the coexistence line and moves the critical point towards smaller pore sizes and higher electrolyte concentrations.

We finally consider the experimental case where the surface charge density varies  with the bulk  $p$H by introducing a charge regulation mechanism. The first-order transition survives and qualitatively corresponds to experimental results for conductivity fluctuations measurements in nanopores~\cite{lev,pasternak}. Moreover, \textcolor{black}{our charge regulation model, at an effective  mean-field level, predicts a jump of the surface charge density as a funciton of $p$H at the transition, which is the signature of a sudden release of hydronium ions from the pore in going from the ionic Vapor to Liquid phase.}

Of course, our model has several limitations. First of all, due to the homogeneous form of the variational screening length, the dielectric/solvation deficit repulsion is slightly underestimated. One interesting possible extension consists in introducing a piecewise screening length, as in Ref.~\cite{hatlo}, but for an initial investigation of the type presented here the technical complications would outweigh the benefits. The model neglects also ion size, which becomes important for bulk systems at high concentration or strongly charged interfaces where ion concentrations may exceed the close packing value. For a cylindrical pore characterized by strong repulsive image forces where ionic concentrations are always far below their bulk value, we expect the ion size effect to play a less important role than in the bulk. However direct correlations, such as ion pairing,  between ``dressed'' ions in the nanopore are neglected and could be taken into account by extending the variational approach adopted in the present work to a second order cumulant expansion. Further investigation will be necessary to estimate the higher order contributions and we expect that it might  change the parameter values at coexistence without modifying our qualitative conclusions.

Our present model also neglects ionic polarizability and treats the water solvent as a dielectric continuum of the same polarizability as in the bulk. It is important to note that due to various effects, the dielectric permittivity within the pore can be lower or higher than that of the bulk~\cite{Marti}. The consideration of this additional complication requires a more detailed solvent model. We are currently working on the generalization of our model to include the polarizability of charged ions, which should yield a more complete picture of the behaviour of large anions in confined geometries. The application of the present model to membrane nanofiltration with practical applications and direct comparison with experiments will be presented in a future work.

\acknowledgments{We thank B. Coasne for helpful discussions. This work was supported in part by the French ANR (project SIMONANOMEM No. ANR-07-NANO-055).}

\appendix
\section{Green's function in cylindrical coordinates : single cylinder}
\label{appendix}

In this appendix, we derive the Green's function $v_0(\br,\br')$ in the presence of a cylindrical dielectric interface of infinite length and radius $a$. The system is characterized by an electric discontinuity defined by $\epsilon(r)=\epsilon_<\Theta(a-r)+\epsilon_>\Theta(r-a)$ and a Debye constant $\kappa(r)=\kappa_<\Theta(a-r)+\kappa_>\Theta(r-a)$, where $r$ denotes the radial distance and $\gtrless$ means $r\gtrless a$. We solve the Debye-H\"{u}ckel equation
\be\label{DHCYbis}
\left[-\nabla(\epsilon(r)\nabla)+\epsilon(r)\kappa^2(r)\right]v_0(\br,\br')=\beta e^2\delta(\br-\br')
\ee
by exploiting the cylindrical symmetry of the system, i.e.
$v_0(\br,\br')=v_0(z-z',\theta-\theta',r,r')$, where $\theta$ stands for the azimuthal angle. To this end, we expand $v_0(\br,\br')$ in Fourier space
\be
v_0(\br,\br')=\sum_{m=-\infty}^{+\infty}e^{im(\theta-\theta')}
\int_{-\infty}^{+\infty}\frac{\mathrm{d}k}{4\pi^2}\;e^{ik(z-z')}\tilde{v}_0(r,r';m,k).
\ee
By injecting this expansion into Eq.~(\ref{DHCYbis}) and using
\bea
\delta(\br-\br')&=&\frac{1}{r}\delta(r-r')\delta(z-z')\delta(\theta-\theta')\nonumber\\
\delta(z-z')&=&\frac{1}{2\pi}\int_{-\infty}^{\infty}\mathrm{d}k\;e^{ik(z-z')}\nonumber\\
\delta(\theta-\theta')&=&\frac{1}{2\pi}\sum_{m=-\infty}^{+\infty}e^{im(\theta-\theta')},
\eea
Eq.~(\ref{DHCYbis}) reduces to
\be\label{DHCLFourier}
\frac{\partial^2\tilde{v}_0}{\partial r^2}+\frac1{r}\frac{\partial\tilde{v}_0}{\partial r}-\left[\varkappa_{\gtrless}^2+\frac{m^2}{r^2}\right]\tilde{v}_0=-\frac{4\pi \ell_\gtrless}{r}\delta(r-r'),
\ee
where we have defined $\varkappa_{\gtrless}^2=k^2+\kappa_{\gtrless}^2$ and $\ell_\gtrless=\beta e^2/(4\pi\epsilon_{\gtrless})$. The homogeneous solutions of Eq.~(\ref{DHCLFourier}) are the modified Bessel functions, $\tilde{v}_0(r,r';m,k)=A(r')I_m(\varkappa r)+B(r')K_m(\varkappa r)$. By taking into account the finiteness of the Green's function on the cylinder axis $r=0$ and at infinity $r\to\infty$ together with the continuity conditions
\bea
\tilde{v}^>_0(r=a)&=&\tilde{v}^<_0(r=a)\\
\left.\epsilon_{\rm m}\frac{\partial\tilde{v}^>_0}{\partial r}\right|_{r=a} &=&
\left.\epsilon_{\rm w}\frac{\partial\tilde{v}^<_0}{\partial r}\right|_{r=a}\\
\left.\frac{\partial\tilde{v}^>_0}{\partial r}\right|_{r=r'}&=&\left.\frac{\partial\tilde{v}^<_0}{\partial r}\right|_{r=r'}-\frac{4\pi \ell_B}{r'},
\eea
one obtains the solution of Eq.~(\ref{DHCYbis}) in the form
\be
v_0^\gtrless(\br,\br')=\ell_\gtrless\frac{e^{-\kappa_\gtrless|\br-\br'|}}{|\br-\br'|}+\delta v_0^\gtrless(\br,\br').
\ee
\begin{widetext}
The first term in the rhs is the usual DH potential in the bulk and the second term, which incorporates the dielectric discontinuity at the interface, reads for $r'<r<a$
\bea
\delta v_0^<(\br,\br')&=&\frac{4\ell_<}{\pi}\sum'_{m\geq0}\cos[m(\theta-\theta')]\int_0^\infty\mathrm{d}k\cos[k(z-z')] F_m(k;\kappa_<,\kappa_>)I_m(\varkappa_<r)I_m(\varkappa_<r')\label{GRCYin}\\
F_m(k;\kappa_<,\kappa_>)&=&\frac{\epsilon_<\varkappa_<K_m(\varkappa_>a)K'_m(\varkappa_<a)-\epsilon_>\varkappa_> K_m(\varkappa_<a)K'_m(\varkappa_>a)}
{\epsilon_>\varkappa_> I_m(\varkappa_<a)K'_m(\varkappa_>a)-\epsilon_<\varkappa_<K_m(\varkappa_>a)I'_m(\varkappa_<a)}.\label{Fm}
\eea
and the prime on the summation sign means that the term $m=0$ must be multiplied by 1/2. Let us note that in the mid-point approximation ($r=0$), the potential reduces to
\be
\delta v_0^<(r=r'=0,z=z',\theta=\theta')=\frac{4\ell_<}{\pi}\int_0^\infty\mathrm{d}k
\frac{\epsilon_<\varkappa_<K_0(\varkappa_>a)K_1(\varkappa_<a)-\epsilon_>\varkappa_> K_0(\varkappa_<a)K_1(\varkappa_>a)}
{\epsilon_<\varkappa_<K_0(\varkappa_>a)I_1(\varkappa_<a)+\epsilon_>\varkappa_> I_0(\varkappa_<a)K_1(\varkappa_>a)}.
\ee
If the source is located outside the cylinder, one obtains in the region $a<r<r'$
\bea
\delta v_0^>(\br,\br')&=&\frac{4\ell_>}{\pi}\sum'_{m\geq0}\cos[m(\theta-\theta')]\int_0^\infty \mathrm{d}k\cos[k(z-z')] G_m(k;\kappa_<,\kappa_>)K_m(q_>r)K_m(q_>r')\\
G_m(k;\kappa_<,\kappa_>)&=&\frac{\epsilon_>q_>I_m(q_<a)I'_m(q_>a)-\epsilon_<q_< I_m(q_>a)I'_m(q_<a)}
{\epsilon_<q_< K_m(q_>a)I'_m(q_<a)-\epsilon_>q_>I_m(q_<a)K'_m(q_>a)}.
\eea

We also report the following coefficients introduced in Eq.~(\ref{FGII})
\bea\label{Jm}
J_0(k;\kappa)&=&I_0^2(\varkappa a)-I_1^2(\varkappa a)\hspace{2mm},\nonumber\\
J_m(k;\kappa)&=&I_m^2(\varkappa a)-I_{m+1}(\varkappa a)I_{m-1}(\varkappa a)\hspace{1cm}\mbox{for}\hspace{2mm}m>0.
\eea

Finally, in the calculation of the variational grand potential Eq.~(\ref{omega}), we perform analytically the integral over the radial distance $r=|\br|$ for the dielectric part which depends on $\delta v_0$,   which simplifies to
\be\label{FGII}
\frac{\Omega_d}{V_{\rm p}} = \frac{\kappa_v^2}{2\pi^2}\sum'_{m\geq0}\int_0^1 \mathrm{d}\xi\int_0^\infty \mathrm{d}k\left[F_m(k;\kappa_v\sqrt\xi,\kappa_>=0)J_m(k;\kappa_v\sqrt\xi)-F_m(k;\kappa_v,\kappa_>=0)J_m(k;\kappa_v)\right].
\ee

\section{Green's function in cylindrical coordinates : concentric cylinders}
\label{appendix}

We report in this appendix the  Green's function solution of Eq.~(\ref{DHCLFourier}), for concentric cylinders of radius $a_1$ and $a_2$ with $a_1<a_2$. We briefly explain in the end the computation of the variational free energy for this more complicated case. The system is characterized by a dielectric discontinuity defined by $\epsilon(r)=\epsilon_m\Theta(a_1-r)+\epsilon_w\Theta(r-a_1)\Theta(a_2-r)+\epsilon_m\Theta(r-a_2)$ and a Debye constant $\kappa(r)=\kappa_v\Theta(r-a_1)\Theta(a_2-r)$ (we consider only the case of ions confined between the two cylinders). The solution of Eq. (\ref{DHCLFourier}) satisfying the boundary conditions
\bea
\tilde{v}^>_0(r=a_1)&=&\tilde{v}^<_0(r=a_1)\\
\tilde{v}^>_0(r=a_2)&=&\tilde{v}^<_0(r=a_2)\\
\left.\epsilon_{\rm w}\frac{\partial\tilde{v}^>_0}{\partial r}\right|_{r=a_1} &=&
\left.\epsilon_{\rm m}\frac{\partial\tilde{v}^<_0}{\partial r}\right|_{r=a_1}\\
\left.\epsilon_{\rm m}\frac{\partial\tilde{v}^>_0}{\partial r}\right|_{r=a_2} &=&
\left.\epsilon_{\rm w}\frac{\partial\tilde{v}^<_0}{\partial r}\right|_{r=a_2}\\
\left.\frac{\partial\tilde{v}^>_0}{\partial r}\right|_{r=r'}&=&\left.\frac{\partial\tilde{v}^<_0}{\partial r}\right|_{r=r'}-\frac{4\pi \ell_B}{r'}
\eea
is given by
\be
v_0(\br,\br')=\ell_B\frac{e^{-\kappa_v|\br-\br'|}}{|\br-\br'|}+\delta v_0(\br,\br')
\ee
with
\bea
\delta v_0(\br,\br')&=&\frac{4\ell_B}{\pi}\sum'_{m\geq0}\cos[m(\theta-\theta')]\int_0^\infty\mathrm{d}k\cos[k(z-z')] \left\{A_m(k;\kappa_v)I_m(\varkappa_vr)I_m(\varkappa_vr')\right.\\
&&\left.+B_m(k;\kappa_v)K_m(\varkappa_vr)K_m(\varkappa_vr')+C_m(k;\kappa_v)\left[I_m(\varkappa_vr)K_m(\varkappa_vr')+I_m(\varkappa_vr')K_m(\varkappa_vr)\right]\right\}\label{GRCYco}\nonumber
\eea
where we have defined
\bea
A_m(k;\kappa_v)&=&\frac{a_m(k;\kappa_v)c_m(k;\kappa_v)}{a_m(k;\kappa_v)b_m(k;\kappa_v)-c_m(k;\kappa_v)d_m(k;\kappa_v)}\nonumber\\
B_m(k;\kappa_v)&=&\frac{d_m(k;\kappa_v)b_m(k;\kappa_v)}{a_m(k;\kappa_v)b_m(k;\kappa_v)-c_m(k;\kappa_v)d_m(k;\kappa_v)}\nonumber\\
C_m(k;\kappa_v)&=&\frac{c_m(k;\kappa_v)d_m(k;\kappa_v)}{a_m(k;\kappa_v)b_m(k;\kappa_v)-c_m(k;\kappa_v)d_m(k;\kappa_v)},\nonumber
\label{ABCm}
\eea
with the coefficients $a_m(k;\kappa_v)$, $bm(k;\kappa_v)$ and $c_m(k;\kappa_v)$ defined according to
\bea
a_m(k;\kappa_v)&=&\epsilon_mkK_m(\varkappa_va_1)I'_m(ka_1)-\epsilon_w\varkappa_v I_m(ka_1)K'_m(\varkappa_va_1)\nonumber\\
b_m(k;\kappa_v)&=&\epsilon_mkI_m(\varkappa_va_2)K'_m(ka_2)-\epsilon_w\varkappa_v K_m(ka_2)I'_m(\varkappa_va_2)\nonumber\\
c_m(k;\kappa_v)&=&\epsilon_w\varkappa_v K_m(ka_2)K'_m(\varkappa_va_2)-\epsilon_mkK_m(ka_2)K'_m(\varkappa_va_2)\nonumber\\
d_m(k;\kappa_v)&=&\epsilon_w\varkappa_v I_m(ka_1)I'_m(\varkappa_va_1)-\epsilon_mkI_m(ka_1)I'_m(\varkappa_va_1),\nonumber
\eea
and $\varkappa_v^2=\kappa_v^2+k^2$. The Green's function evaluated at the same point is given by
\bea\label{GRCYcoII}
v_0(\br,\br')=v_c^b(0,0)-\kappa_v\ell_B+\delta v_0(\br,\br')
\eea
with
\be
\delta v_0(\br,\br)=\frac{4\ell_B}{\pi}\sum'_{m\geq0}\int_0^\infty\mathrm{d}k \left\{A_m(k;\kappa_v)I_m(\varkappa_vr)^2
+B_m(k;\kappa_v)K_m(\varkappa_vr)^2+2C_m(k;\kappa_v)I_m(\varkappa_vr)K_m(\varkappa_vr)\right\}\label{GRCYcoIII}.
\ee
\end{widetext}

The DH pressure from an hypothetical bulk solution is now given by
\be
p=-\frac{\Omega_v}L=-(a_2^2-a_1^2)\frac{\kappa_v^3}{24}
\ee
and the spatial integrations in Eq.~(\ref{omega}) are evaluated from $r=a_1$ to $r=a_2$. As in the case with a single cylindrical pore, the obtained grand-potential energy $\Omega_v$ is minimized with respect to $\kappa_v$ to find the optimal solution to the variational problem.


\begin{thebibliography}{99}

\bibitem {Holm} C. Holm, P. Kekicheff, and R. Podgornik, Electrostatic Effects in Soft Matter and Biophysics, Kluwer Academic,Dordrecht (2001).
\bibitem {Dub} M. Dubois and T. Zemb, Langmuir, \textbf{7}, 1352 (1991).
\bibitem {Synt} S. D.  Shoemaker and T. K. Vanderlick,  Biophys. J., \textbf{83}, 2007 (2002).
\bibitem {Bont} D. J. Bonthuis et al., Phys. Rev. Lett., \textbf{97}, 128104 (2006).
\bibitem {DLVO} Verwey EJW, Overbeek JThG, Theory of the stability of lyophobic colloids, Elsevier, Amsterdam (1948).
\bibitem {Pars}  A. Parsegian, Nature, \textbf{221} ,844 (1969).
\bibitem {NSDH} D. G. Levitt , Biophys. J., \textbf{22}, 209 (1978).
\bibitem {Levin} Y. Levin, Europhys. Lett., \textbf{76} , 163 (2006).
\bibitem {ShlovsSC} B.I. Shklovskii, Phys. Rev. E, \textbf{60}, 5802 (1999).
\bibitem {NetzSC} A. G. Moreira and R. R. Netz,  Europhys. Lett., \textbf{52} (6), 705 (2000) .
\bibitem {Attard} P. Attard,  D.J. Mitchell , and B.W. Ninham, J. Chem. Phys., \textbf{88}, 4987 (1988); ibid. \textbf{89}, 4358.
\bibitem {Zeks} R. Podgornik and B. Zeks, J. Chem. Soc. Faraday Trans. II \textbf{84} , 611 (1988).
\bibitem {netzCOR} R.R. Netz and H. Orland, Eur. Phys. J. E, \textbf{1}, 203 (2000).
\bibitem {PodKan} M. Kanduc and R. Podgornik, Eur. Phys. J. E, \textbf{23}, 265 (2007).
\bibitem {DeanHor} D.S. Dean and R.R. Horgan,, Phys. Rev. E \textbf{70}, 011101 (2004)
\bibitem {DavidCYL} D.S. Dean and R.R. Horgan, J. Phys. C. \textbf{17}, 3473, (2005).
\bibitem {NajiKand} M. Kanduc, A. Naji and R. Podgornik, J. Chem. Phys., \textbf{132}, 224703 (2010).
\bibitem {netzREV} A. Naji et al., Physica A, \textbf{352}, 131 (2005).
\textcolor{black}{
\bibitem {Arnold1} A. Arnold and C. Holm, Comput. Phys. Commun. \textbf{148}, 327 (2002).
\bibitem {Arnold2} Tyagi S., Arnold A. and C. Holm, J. Chem. Phys. \textbf{127}, 154723 (2007).
\bibitem {Jho1} Y. S. Jho et al., Phys. Rev. E \textbf{76}, 011920 (2007).
\bibitem {Jho2} Y. S. Jho et al.,  J. Chem. Phys. \textbf{129}, 134511 (2008).
\bibitem {Jho3} Y. S. Jho et al., Phys. Rev. Lett. \textbf{101}, 188101 (2008).
}
\bibitem {netz} R.R. Netz and H. Orland, Eur. Phys. J. E, \textbf{11}, 301 (2003).
\bibitem {curtis} R.A. Curtis and L. Lue,. J. Chem. Phys., \textbf{123}, 174702 (2005)
\bibitem {hatlo} M.M. Hatlo, R.A. Curtis and L. Lue, J. Chem. Phys., \textbf{128}, 164717 (2008)
\bibitem {hatlo_review} M.M. Hatlo and L. Lue, Soft Matter, \textbf{4}, 1582 (2008)
\bibitem {Nous} S. Buyukdagli, M. Manghi, and J. Palmeri, Phys. Rev. E \textbf{81}, 041601 (2010).
\bibitem {onsager_samaras} L. Onsager and N. Samaras, J. Chem. Phys. \textbf{2}, 528 (1934).
\bibitem {prl} S. Buyukdagli, M. Manghi, and J. Palmeri, \textcolor{black}{Phys. Rev. Lett., \textbf{105}, 158103 (2010)}.
\bibitem {yarosh} A.E. Yaroshchuk, Adv. Colloid Interf. Sci., \textbf{85}, 193 (2000).
\bibitem {loeb} A. L. Loeb, J. Colloid. Sci.,\textbf{6}, 75 (1950).
\bibitem {Dresner} L. Dresner, Desalination, \textbf{15}, 39 (1974).
\bibitem {lev} A.A. Lev \textit{et al.}, Proc. R. Soc. Lond. B \textbf{252}, 187 (1993).
\bibitem {JANCO} B. Jancovici and X. Artru, Mol. Phys., \textbf{49}, 487 (1983).
\bibitem {note} S. Buyukdagli, M. Manghi, and J. Palmeri, unpublished.
\bibitem {mcquarrie} D.A. McQuarrie, \emph{Statistical Mechanics} chap.15 (University Science Book, New York, 2000).
\bibitem {Lue1} J. Groenewold, J. Chem. Phys., \textbf{107}, 9668 (1997).
\bibitem {DiExp} T. Arawaka and S. N. Timasheff, Biochemistery, \textbf{21}, 6545 (1982).
\bibitem {stell} G. Stell \textit{et al.} Phys. Rev. Lett. \textbf{37}, 1369 (1976).
\bibitem {singh} R.R. Singh and K.S. Pitzer, J. Chem. Phys. \textbf{92}, 6775 (1990).
\bibitem {fisher} M.E. Fisher and Y. Levin, Phys. Rev. Lett. \textbf{71}, 3826 (1993).
\bibitem {hynn} A.-P. Hynninen and A.Z. Panagiotopoulos, Mol. Phys. \textbf{106}, 2038 (2008).
\bibitem {evans} R. Evans, U.M.B. Marconi, P. Tarazona, J. Chem. Phys. \textbf{84}, 2376 (1986).
\bibitem {evans_review} R. Evans, J. Phys.: Condens. Matter \textbf{2}, 8989 (1990).
\bibitem {gubbins} G.S. Heffelfinger, F. van Swol, and K.E. Gubbins, J. Chem. Phys. \textbf{89}, 5202 (1988).
\bibitem {peterson} B.K. Peterson \textit{et al.}, J. Chem. Phys. \textbf{88}, 6487 (1988).
\bibitem {roth} R. Roth and K.M. Kroll,  J. Phys.: Condens. Matter \textbf{18}, 6517 (2006).
\bibitem {beckstein} O. Beckstein and M.S.P. Sansom, Proc. Natl. Acad. Sc. USA \textbf{100}, 7063 (2003).
\bibitem {onsager} L. Onsager, Chem. Rev., \textbf{13}, 73 (1933).
\bibitem {Hille} B. Hille, Ion Channels of Excitable Membranes, Sinauer Associates, Sunderland, MA (2001).
\bibitem {Kamen} J. Zhang, A. Kamenev, B. I. Shklovskii, Phys. Rev. Lett. \textbf{95}, 148101 (2005).
\bibitem {Kameneev} J. Zhang, A. Kamenev, and B. I. Shklovskii, Phys. Rev. E \textbf{73}, 051205 (2006).
\bibitem {pasternak} C.A. Pasternak \textit{et al.}, Colloids Surf. A \textbf{77}, 119 (1993).
\bibitem {hijnen} H.J.M. Hijnen and J.A.M. Smit, Biophys. Chem. \textbf{41}, 101 (1991).
\textcolor{black}{\bibitem {PodChrReg} R. Podgornik, J. Chem. Phys., \textbf{91}, 5840 (1989).}
\textcolor{black}{\bibitem {ninham} B.W. Ninham and V.A. Parsegian, J. Theor. Biol., \textbf{31}, 405 (1971).}
\bibitem {Manciu} M. Manciu and E. Ruckenstein, Advances in Colloid and Interface Science, \textbf{105}, 63 (2003).
\bibitem {Marti} Marti et al., J. Phys. Chem. B \textbf{110}, 23987 (2006)




\end{thebibliography}
\end{document}